\documentclass[%
 aip,
 amsmath,amssymb,
 reprint,%
]{revtex4-1}

\usepackage{amsmath} 
\usepackage{multirow} 
\usepackage{xcolor} 
\usepackage{CJKutf8}

\usepackage{graphicx}
\usepackage{dcolumn}
\usepackage{bm}

\usepackage[utf8]{inputenc}
\usepackage[T1]{fontenc}
\usepackage{mathptmx}
\usepackage{etoolbox}

\makeatletter
\def\@email#1#2{%
 \endgroup
 \patchcmd{\titleblock@produce}
  {\frontmatter@RRAPformat}
  {\frontmatter@RRAPformat{\produce@RRAP{*#1\href{mailto:#2}{#2}}}\frontmatter@RRAPformat}
  {}{}
}%
\makeatother
\begin{document}

\preprint{AIP/123-QED}

\title[Micellar Segment Number Density Underlies Stress-Optical Response of Wormlike Micellar Solutions]{Micellar Segment Number Density Underlies Stress-Optical Response of Wormlike Micellar Solutions}

\author{Masakazu Muto\begin{CJK}{UTF8}{min}(武藤真和)\end{CJK}\textsuperscript{*}}
 \email[Authors to whom correspondence should be addressed: ]{muto.masakazu@nitech.ac.jp and tamano.shinji@nitech.ac.jp}
 \affiliation{ 
 Department of Mechanical Engineering, Nagoya Institute of Technology, Nagoya, Japan. Gokiso Showa-ku, Nagoya-shi, Aichi 466-8555, Japan. 
 }

\author{Naoki Kako \begin{CJK}{UTF8}{min}(加古直樹)\end{CJK}}%
\affiliation{ 
Department of Mechanical Engineering, Nagoya Institute of Technology, Nagoya, Japan. Gokiso Showa-ku, Nagoya-shi, Aichi 466-8555, Japan.
}%

\author{Shinji Tamano \begin{CJK}{UTF8}{min}(玉野真司)\end{CJK}\textsuperscript{*}}
\affiliation{ 
Department of Mechanical Engineering, Nagoya Institute of Technology, Nagoya, Japan. Gokiso Showa-ku, Nagoya-shi, Aichi 466-8555, Japan.
}%

\date{\today}

\begin{abstract}
Wormlike micellar solutions exhibit complex viscoelastic behavior that strongly depends on micellar length and structure. However, the effects of these structural characteristics on the relationship between flow-induced birefringence and stress remain unclear. In this study, the stress-optical response under uniaxial extensional flow was investigated for cetyltrimethylammonium bromide (CTAB)/sodium salicylate (NaSal) aqueous solutions with systematically varied compositions. Full-field extensional rheo-optical data and dynamic viscoelastic measurements were analyzed to evaluate the stress-optical coefficient, intrinsic birefringence, microscopic optical properties, and micellar structural parameters. A nearly linear relationship between birefringence and extensional stress was observed during the filament-thinning process, consistent with the stress-optical rule. The intrinsic birefringence and stress-optical coefficient increased in magnitude with increasing micellar molar concentration, whereas the local optical anisotropy of individual micellar segments remained approximately constant. Although the stress-optical coefficient showed apparent correlations with micellar length and Kuhn length, neither exhibited a clear systematic relationship at a fixed micellar molar concentration. Instead, the magnitude of the stress-optical coefficient was approximately proportional to the geometrical micellar volume fraction, indicating a common empirical relationship across solutions with different micellar lengths and potentially different structural morphologies. These results suggest that the stress-optical response is more closely associated with the number density of micellar segments than with changes in their local optical properties or overall micellar length. This relationship offers a potential approach to estimate the stress-optical coefficient from the composition of a given solution to facilitate non-contact measurements of extensional stress using flow-induced birefringence.
\end{abstract}

\maketitle

\section{\label{sec:1}Introduction}
\begin{figure*}[t]
\centering
\includegraphics[width=0.74\textwidth]{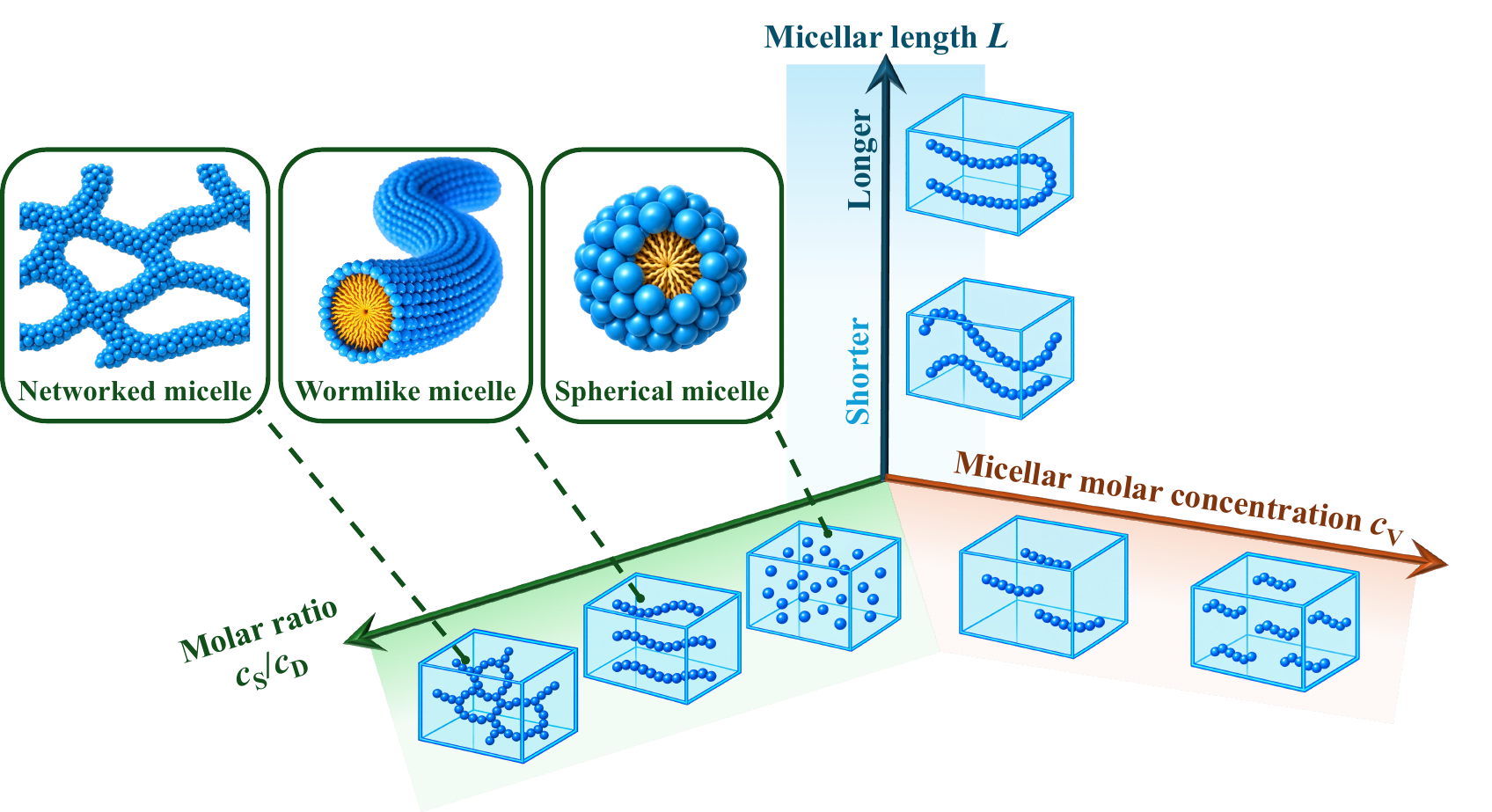}
\caption{
Schematic representation of the relationship among the CTAB/NaSal molar ratio $c_{\mathrm{S}}/c_{\mathrm{D}}$, micellar molar concentration $c_{\mathrm{V}}$, micellar length $L$, and possible micellar morphologies. The illustrated structures conceptually represent relatively short micelles, elongated and entangled wormlike micelles, and branched or networked micelles. Some elements in this figure were generated using ChatGPT (OpenAI, Inc.).}
\label{fig:1}
\end{figure*}

\begin{CJK}{UTF8}{min}
Wormlike micelles formed by the self-assembly of surfactant molecules in aqueous solutions possess flexible structures with a high aspect ratio and exhibit pronounced viscoelasticity through mutual entanglement above a certain concentration.  \cite{liberatore2009microstructure,dreiss2007wormlike,walker2001rheology,lequeux1996structure,cates1990statics,rehage1991viscoelastic,turner1992linear,granek1992stress,turner1991linear,cates1987reptation} Although wormlike micelles exhibit mechanical responses similar to those of polymer chains, the association state of the surfactant molecules changes dynamically, and the micelles repeatedly undergo scission and recombination. Accordingly, wormlike micelles are also referred to as living polymers.\cite{cates1990statics, cates1987reptation}  In addition, their stress relaxation is characterized by motions similar to those of polymer chains as well as scission--recombination processes. \cite{cates1990statics,rehage1991viscoelastic,turner1992linear,granek1992stress,turner1991linear,cates1987reptation} Therefore, the rheological properties of wormlike micellar solutions strongly depend not only on the concentration of the surfactant but also on the internal structure of the wormlike micelles, including their micellar length, flexibility, entanglement, branching, and cross-linking. \cite{liberatore2009microstructure,dreiss2007wormlike,walker2001rheology,lequeux1996structure,cates1990statics,rehage1991viscoelastic,turner1992linear,granek1992stress,turner1991linear} In particular, cetyltrimethylammonium bromide (CTAB)/sodium salicylate (NaSal) aqueous solutions have been widely studied as representative model fluids because they form long wormlike micelles and exhibit pronounced viscoelasticity even at relatively low surfactant concentrations. \cite{shikata1987micelle,shikata1989micelle,shikata1988micelle,inoue2005nonlinear,lin2001influence,kadoma1997structural,shikata1988nonlinear} In these solutions, the micellar length, entanglement state, and branched or cross-linked structures change depending on the concentrations of CTAB and NaSal. \cite{shikata1989micelle,shikata1988micelle,inoue2005nonlinear,lin2001influence,kadoma1997structural,shikata1988nonlinear} In particular, the CTAB/NaSal molar ratio $c_{\mathrm{S}}/c_{\mathrm{D}}$ [-] is a compositional parameter associated with the structural transition from wormlike to networked micelles. \cite{shikata1988micelle,inoue2005nonlinear,lin2001influence,kadoma1997structural} This value is defined as the ratio of the NaSal molar concentration $c_{\mathrm{S}}$ [M] to the CTAB molar concentration $c_{\mathrm{D}}$ [M]. In contrast, the micellar molar concentration $c_{\mathrm{V}}$ [M] represents the amount of micellar units ($\mathrm{CTA}^{+}\mathrm{Sal}^{-}$ complexes) per unit volume, whereas the micellar length $L$ [m] represents the average length of a population of wormlike micelles exhibiting a given distribution of lengths. Therefore, the number of wormlike micelles per unit volume may differ with different $L$ even at the same $c_{\mathrm{V}}$. Thus, the parameters $c_{\mathrm{S}}/c_{\mathrm{D}}$, $c_{\mathrm{V}}$, and $L$, respectively, denote the solution composition related to their structural morphology, the amount of micellar units, and the micellar length to characterize different aspects of wormlike micelles. 
The relationships between these parameters and the internal structures are schematically illustrated in Fig.~\ref{fig:1}.
Clarifying the relationship between the internal structure and flow properties provides some fundamental insights into self-assembled systems. These findings also support new guidance for selecting solution compositions to obtain desired viscosity and relaxation properties as well as for designing materials while considering structural changes during flow.

Rheo-optics combines rheological and birefringence measurements. This approach has been applied to evaluate the internal structures of complex fluids through flow-induced birefringence, particularly the deformation and orientation of polymer chains and wormlike micelles under flow. \cite{chow1984response,worby2024examination,sato2024twodimensional,yamamoto2024elucidating,ito2016shearinduced,ito2015temporal,meyer1993investigation,lodge1955variation,philippoff1957flow,humbert1998stress,humbert1998flow,wheeler1996structure,shikata1994rheooptical,decruppe2003flow,janeschitz1983polymer, sridhar2000birefringence, talbott1979streaming,haward2011extensional} The optical anisotropy of polymer chains or wormlike micelles appears as macroscopic birefringence when they become oriented by flow. Therefore, simultaneous measurements of stress and birefringence can be used to relate the mechanical response of a fluid to changes in its internal structure. Stress-optical responses have been investigated extensively in complex fluids such as polymer and wormlike micellar solutions under shear flow along with the orientation and deformation of internal structures. \cite{chow1984response,worby2024examination,sato2024twodimensional,yamamoto2024elucidating,ito2016shearinduced,ito2015temporal,meyer1993investigation,lodge1955variation,philippoff1957flow,humbert1998stress,humbert1998flow,wheeler1996structure,shikata1994rheooptical,decruppe2003flow} The orientation of polymer chains and wormlike micelles under uniaxial and planar extensional flows has also been evaluated using birefringence. \cite{sridhar2000birefringence,talbott1979streaming,haward2011extensional,rothstein2002comparison} More recently, the use of high-speed polarization cameras has enabled the two-dimensional distributions of birefringence and optical principal-axis orientation to be obtained with high temporal and spatial resolutions, which allows the relationship between stress responses and internal structures under flow to be evaluated in greater detail. \cite{worby2024examination,sato2024twodimensional,yamamoto2024elucidating,araida2025analysis} 

Birefringence is an optical quantity that reflects the internal structure of a fluid and can be quantitatively related to stress under specific conditions. When the optical principal axes coincide with the principal stress axes, the birefringence $\delta n$ [-] is proportional to the principal stress difference under conditions where the stress-optical rule (SOR) is valid: \cite{lodge1955variation,philippoff1957flow,humbert1998stress,humbert1998flow,wheeler1996structure,shikata1994rheooptical,decruppe2003flow,sridhar2000birefringence, janeschitz1983polymer}
\begin{equation}
\delta n = n_{\parallel}-n_{\perp}
= C(\sigma_{\parallel}-\sigma_{\perp}),
\label{eq:1}
\end{equation}
where $n_{\parallel}$ [-] and $n_{\perp}$ [-] are the refractive indices along the optical principal axes, $\sigma_{\parallel}$ [Pa] and $\sigma_{\perp}$ [Pa] are the corresponding principal stresses, and $C$ [$\mathrm{Pa}^{-1}$] is the stress-optical coefficient. When $C$ is known, the principal stress difference ($\sigma_{\parallel} - \sigma_{\perp}$) can be estimated from birefringence measurements. Given that $C$ characterizes the relationship between the internal structure of complex fluids and their optical response to stress, clarifying its dependence on solution composition and microstructure is essential to quantitatively evaluate stress based on birefringence. Indeed, the first author has previously contributed to the development of birefringence-based stress measurement techniques using high-speed polarization imaging, including stress evaluation in three-dimensional flows of cellulose nanocrystal suspensions \cite{nakamine2024flow} and the visualization of dynamic stress interactions between solids and liquids. \cite{muto2025solidliquid}

Building on these developments, the authors developed a full-field extensional rheo-optical measurement method \cite{muto2025fullfield} that combines a liquid dripping method \cite{muto2026scale,tamano2017dynamics} with a high-speed polarization camera to improve the accuracy of stress-optical coefficient $C$ measurements. Using this method, the extensional stress and two-dimensional birefringence field of CTAB/NaSal aqueous solutions under uniaxial extensional flow were measured simultaneously and without contact. An approximately linear relationship between birefringence and extensional stress accompanied by micellar alignment along the extensional direction was observed in the elasto-capillary (EC) regime in which elastic stress balances capillary forces. Furthermore, $C$ was approximately proportional to the micellar molar concentration $c_{\mathrm{V}}$, whereas no clear dependence on the CTAB/NaSal molar ratio $c_{\mathrm{S}}/c_{\mathrm{D}}$ was observed. \cite{muto2025fullfield}
However, changes in $c_{\mathrm{V}}$ and $c_{\mathrm{S}}/c_{\mathrm{D}}$ also affect micellar structure, including micellar length, entanglement, branching, and cross-linking. \cite{shikata1989micelle,shikata1988micelle,inoue2005nonlinear,lin2001influence,kadoma1997structural,shikata1988nonlinear} In particular, the micellar length $L$ influences solution viscosity and stress relaxation and may affect micellar orientation, extension, and flow-induced scission under uniaxial extensional flow. \cite{liberatore2009microstructure,dreiss2007wormlike,walker2001rheology,lequeux1996structure,cates1990statics,rehage1991viscoelastic,turner1992linear,granek1992stress,turner1991linear,cates1987reptation,kato2002steady,pathak2006rheooptics,rothstein2003transient,haward2012extensional,fischer1997branched,omidvar2019detecting} Furthermore, entangled and branched or cross-linked micellar networks may exhibit different stress-generation and stress-relaxation mechanisms. \cite{inoue2005nonlinear,lin2001influence,kadoma1997structural,fischer1997branched,omidvar2019detecting} Therefore, the relationship between the stress-optical coefficient $C$ and micellar length $L$ must be clarified further to understand how changes in micellar structure contribute to the dependence of $C$ on concentration observed in the previous study.\cite{muto2025fullfield}

Based on these considerations, the objective of this study is to clarify the relationship between the micellar length $L$ and the stress-optical response under uniaxial extensional flow for CTAB/NaSal aqueous solutions with systematically varying CTAB/NaSal molar ratios $c_{\mathrm{S}}/c_{\mathrm{D}}$ and micellar molar concentrations $c_{\mathrm{V}}$. First, the micellar length $L$ was estimated from dynamic viscoelastic measurements using a rotational rheometer, and the validity of the calculation method was examined by comparison with values reported in previous studies. \cite{granek1992stress} Next, the polarizability anisotropies of micellar segments and equivalent Kuhn segments were calculated from the intrinsic birefringence and stress-optical coefficient using the full-field extensional rheo-optical measurement data obtained in the previous study. \cite{muto2025fullfield} This extends the evaluation of the macroscopic stress-optical response to that of microscopic optical anisotropy at different structural scales. Furthermore, the effects of the micellar molar concentration, CTAB/NaSal molar ratio, micellar length, and Kuhn length on the stress-optical response were compared by evaluating the relationships between these optical properties and the micellar length and Kuhn length, and the structural factors governing the stress-optical response were examined. In addition, we established a common empirical relationship for CTAB/NaSal aqueous solutions with different internal structures by organizing the stress-optical response based on the micellar volume fraction.

\section{\label{sec:2}Solution properties}
\subsection{\label{sec:2.1}Sample solutions}
We investigated the same 17 CTAB/NaSal aqueous solutions as in our previous study \cite{muto2025fullfield} with varying molar concentrations of CTAB and NaSal. The CTAB molar concentration $c_{\mathrm{D}}$, NaSal molar concentration $c_{\mathrm{S}}$, micellar molar concentration $c_{\mathrm{V}}$, CTAB/NaSal molar ratio $c_{\mathrm{S}}/c_{\mathrm{D}}$, refractive index $n$ [-], and shear viscosity $\eta_{0.1}$ [$\mathrm{Pa\,s}$] at the minimum shear rate of $\dot{\gamma} = 0.1\,\mathrm{s}^{-1}$ for each solution are listed in Table~\ref{tab:1}. 
\begin{table}[tb!]
    \centering
    \caption{Compositions and physical properties of the CTAB/NaSal aqueous solutions
used in this study. Here, $c_{\mathrm{D}}$, $c_{\mathrm{S}}$, and
$c_{\mathrm{V}}$, respectively, denote the CTAB molar concentration,
NaSal molar concentration, and micellar molar concentration.
The ratio $c_{\mathrm{S}}/c_{\mathrm{D}}$ denotes the CTAB/NaSal
molar ratio, and $n$ denotes the refractive index. The viscosity
$\eta_{0.1}$ is the mean of six measurements obtained at the minimum
shear rate of $\dot{\gamma} = 0.1\,\mathrm{s}^{-1}$ and is reported
to three significant figures.}
    \label{tab:1}
    \setlength{\tabcolsep}{3pt}
    \renewcommand{\arraystretch}{1.35}
    \begin{tabular*}{\columnwidth}{@{\extracolsep{\fill}}lcccccc@{}}
    \hline\hline
    Solution
    & $c_{\mathrm{D}}$
    & $c_{\mathrm{S}}$
    & $c_{\mathrm{V}}~(=c_{\mathrm{D}})$
    & $c_{\mathrm{S}}/c_{\mathrm{D}}$
    & $n$
    & $\eta_{0.1}$ \\
    No.
    & [M]
    & [M]
    & [M]
    & [-]
    & [-]
    & [$\mathrm{mPa\,s}$] \\
    \hline
    $\mathrm{I}$
    & 0.005 & 0.038 & 0.005
    & \multirow{5}{*}{7.7} & 1.335 & 95.0 \\
    $\mathrm{I\hspace{-0.6pt}I}$
    & 0.010 & 0.077 & 0.010 & & 1.336 & 1140 \\
    $\mathrm{I\hspace{-0.6pt}I\hspace{-0.6pt}I}$
    & 0.015 & 0.115 & 0.015 & & 1.338 & 2540 \\
    $\mathrm{I\hspace{-0.6pt}V}$
    & 0.020 & 0.153 & 0.020 & & 1.339 & 5250 \\
    $\mathrm{V}$
    & 0.025 & 0.192 & 0.025 & & 1.341 & 7590 \\
    \hline
    $\mathrm{V\hspace{-0.6pt}I}$
    & 0.030 & 0.230 & 0.030 & 7.7 & 1.342 & 4700 \\
    \hline
    $\mathrm{V\hspace{-0.6pt}I\hspace{-0.6pt}I}$
    & \multirow{11}{*}{0.030}
    & 0.030
    & \multirow{11}{*}{0.030}
    & 1.0 & 1.336 & 12400 \\
    $\mathrm{V\hspace{-0.6pt}I\hspace{-0.6pt}I\hspace{-0.6pt}I}$
    & & 0.060 & & 2.0 & 1.337 & 5270 \\
    $\mathrm{I\hspace{-0.6pt}X}$
    & & 0.090 & & 3.0 & 1.338 & 7590 \\
    $\mathrm{X}$
    & & 0.120 & & 4.0 & 1.339 & 10600 \\
    $\mathrm{X\hspace{-0.6pt}I}$
    & & 0.150 & & 5.0 & 1.340 & 12000 \\
    $\mathrm{X\hspace{-0.6pt}I\hspace{-0.6pt}I}$
    & & 0.180 & & 6.0 & 1.341 & 11900 \\
    $\mathrm{X\hspace{-0.6pt}I\hspace{-0.6pt}I\hspace{-0.6pt}I}$
    & & 0.210 & & 7.0 & 1.342 & 12000 \\
    $\mathrm{X\hspace{-0.6pt}I\hspace{-0.6pt}V}$
    & & 0.270 & & 9.0 & 1.344 & 3530 \\
    $\mathrm{X\hspace{-0.6pt}V}$
    & & 0.300 & & 10.0 & 1.345 & 2860 \\
    $\mathrm{X\hspace{-0.6pt}V\hspace{-0.6pt}I}$
    & & 0.330 & & 11.0 & 1.346 & 1300 \\
    $\mathrm{X\hspace{-0.6pt}V\hspace{-0.6pt}I\hspace{-0.6pt}I}$
    & & 0.360 & & 12.0 & 1.347 & 709 \\
    \hline\hline
    \end{tabular*}
\end{table}
The viscosity $\eta_{0.1}$ was calculated as the mean of six shear-viscosity measurements obtained under identical conditions. The dependence of the refractive index and shear viscosity on the solution composition is described in Secs.~\ref{sec:2.2} and \ref{sec:2.3}, respectively.

The micellar molar concentration $c_{\mathrm{V}}$ listed in Table~\ref{tab:1} is defined as follows. In CTAB/NaSal aqueous solutions, the $\mathrm{CTA}^{+}\mathrm{Sal}^{-}$ complex assumed to form through the 1:1 association of $\mathrm{CTA}^{+}$ derived from CTAB and $\mathrm{Sal}^{-}$ derived from NaSal is treated as a micellar unit constituting wormlike micelles. \cite{shikata1987micelle,shikata1989micelle,shikata1988micelle} In this study, the NaSal molar concentration was equal to or greater than that of CTAB ($c_{\mathrm{S}}/c_{\mathrm{D}} \geq 1$). Assuming that all CTAB molecules contribute to the formation of micellar units, we defined the micellar molar concentration as $c_{\mathrm{V}} = c_{\mathrm{D}}$, following our previous study.\cite{muto2025fullfield} Thus, $c_{\mathrm{V}}$ [M] represents the molar concentration of micellar units, which corresponds to a number density of $10^{3} c_{\mathrm{V}} N_{\mathrm{A}}$ [$\mathrm{m}^{-3}$], where $N_{\mathrm{A}}$ [$\mathrm{mol}^{-1}$] is the Avogadro constant. The factor $10^{3}$ accounts for the unit conversion of $c_{\mathrm{V}}$ from M to $\mathrm{mol~m^{-3}}$. Importantly, $c_{\mathrm{V}}$ does not represent the molar concentration of individual wormlike micelles. Even at the same $c_{\mathrm{V}}$, the number of micellar units constituting an individual wormlike micelle may vary with the micellar length $L$; consequently, the number density of wormlike micelles per unit volume may also differ.

\begin{figure}[t!]
\centering
\includegraphics[width=0.88\columnwidth]{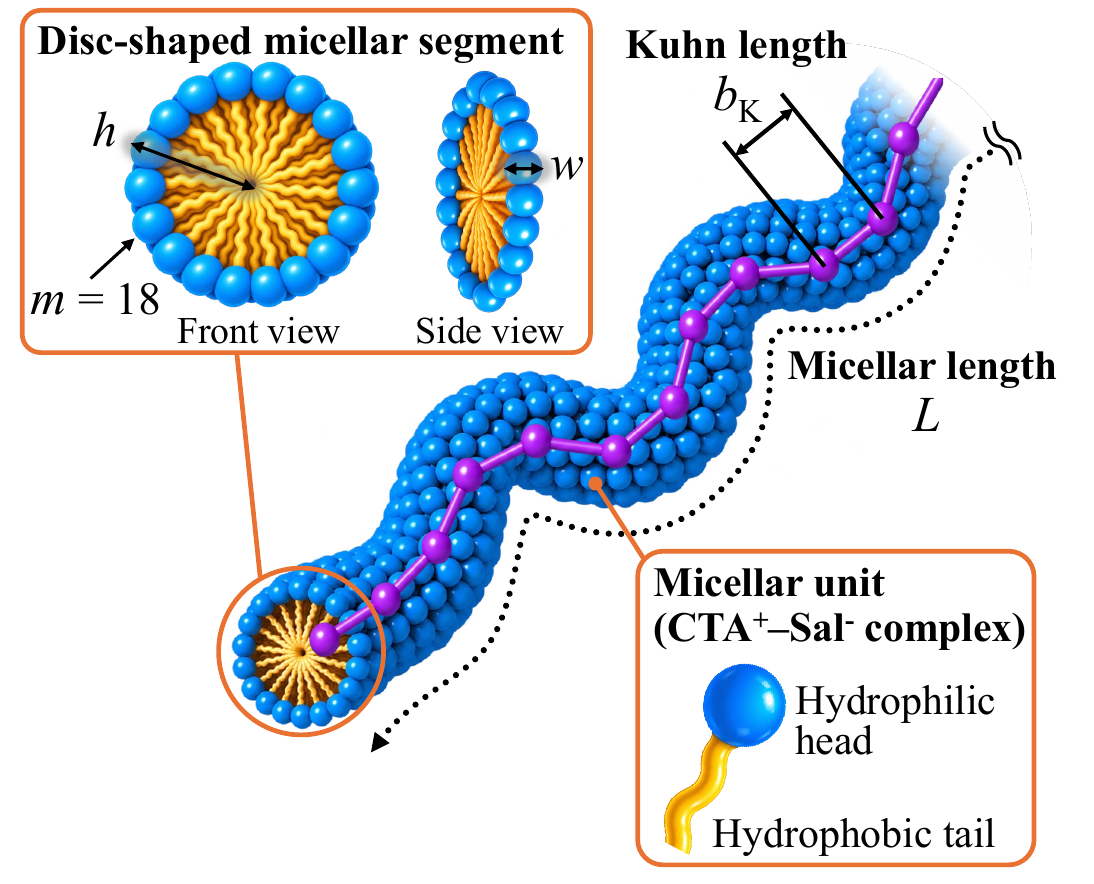}
\caption{
Schematic of the hierarchical structure of a wormlike micelle in a CTAB/NaSal aqueous solution based on the model proposed by Shikata et al. \cite{shikata1994rheooptical} The lower-right inset shows a micellar unit consisting of a 1:1 $\mathrm{CTA}^{+}\mathrm{Sal}^{-}$ complex, while the upper-left inset shows front and side views of a disk-shaped micellar segment composed of $m = 18$ units. The hydrophobic tails face inward, whereas the hydrophilic heads face outward. The segment radius and thickness are denoted by $h$ and $w$, respectively. Successive segments form a wormlike micelle with a contour length $L$ (Fig.~\ref{fig:1}). The magenta line represents the equivalent freely jointed chain model of the wormlike micelle,\cite{Vogtt2015} with $b_{\mathrm{K}}$ denoting the Kuhn segment length. Some elements in this figure were generated using ChatGPT (OpenAI, Inc.).}
\label{fig:2}
\end{figure}

Shikata et al. \cite{shikata1994rheooptical} modeled CTAB/NaSal wormlike micelles as structures composed of disk-shaped micellar segments connected along the axial direction, with each segment consisting of $m$ micellar units. They estimated a value of $m = 18$ based on a geometrical evaluation. Hereafter, this structural model is referred to as the micellar segment chain model. A schematic of a wormlike micelle based on this model is shown in Fig.~\ref{fig:2}.
As shown in the figure, a micellar unit corresponds to a $\mathrm{CTA}^{+}\mathrm{Sal}^{-}$ complex, and $m$ micellar units assemble to form one disk-shaped micellar segment. The successive segments then form a wormlike micelle. In this study, following previous works by Shikata et al., \cite{shikata1994rheooptical} we set $m = 18$ and evaluated the number density of micellar segments per unit volume as $10^{3}c_{\mathrm{V}}N_{\mathrm{A}}/m$ [$\mathrm{m}^{-3}$]. This number density is proportional to $c_{\mathrm{V}}$ because $m$ is assumed to be constant in this model.
In contrast, the statistical properties of wormlike micelles as flexible chains can be represented by an equivalent freely jointed chain model.\cite{Vogtt2015} In this model,\cite{Vogtt2015} a continuously bending wormlike micelle is approximated as a chain of rigid segments that can rotate freely relative to one another. A virtual rigid segment that forms an equivalent chain with the same contour length and mean-square end-to-end distance as an actual wormlike micelle is referred to as an equivalent Kuhn segment,\cite{yamakawa1971modern} and its length is defined as the Kuhn length $b_{\mathrm{K}}$ [m]. Fig.~\ref{fig:2} shows both a wormlike micelle composed of micellar segments and its corresponding equivalent Kuhn chain. Whereas a micellar segment represents a local structural unit of a micelle, an equivalent Kuhn segment is a statistical unit that represents the flexibility of the chain. These two structural units are therefore distinguished from each other.

The 17 sample solutions used in this study were designed to evaluate the changes in the stress-optical response associated with $c_{\mathrm{V}}$ and $c_{\mathrm{S}}/c_{\mathrm{D}}$ separately. For Solutions I--VI, the CTAB/NaSal molar ratio was fixed at $c_{\mathrm{S}}/c_{\mathrm{D}} = 7.7$, while $c_{\mathrm{V}}$ was varied from $0.005$ to $0.030\,\mathrm{M}$. These samples were used to evaluate the stress-optical response associated with changes in the micellar molar concentration while keeping the CTAB/NaSal molar ratio constant. For Solutions VII--XVII, the micellar molar concentration was fixed at $c_{\mathrm{V}} = 0.030\,\mathrm{M}$, while the CTAB/NaSal molar ratio was varied from $c_{\mathrm{S}}/c_{\mathrm{D}} = 1.0$ to $12$. In CTAB/NaSal aqueous solutions, micelles grow with increasing NaSal concentration, and their morphology may change from entangled wormlike micelles to networked micelles with branching and cross-linking depending on the conditions. \cite{shikata1988micelle,inoue2005nonlinear,lin2001influence,kadoma1997structural,muto2025fullfield} Therefore, Solutions VII--XVII were used to evaluate the relationship between the micellar length and structural morphology associated with changes in the CTAB/NaSal molar ratio $c_{\mathrm{S}}/c_{\mathrm{D}}$ and the stress-optical response while keeping the micellar molar concentration $c_{\mathrm{V}}$ constant.

\subsection{\label{sec:2.2}Refractive index}
The refractive index $n$ of the solution is required to evaluate the polarizability anisotropies\cite{shikata1994rheooptical} $\Delta\alpha$ and $\Delta\alpha_0$ from the measured birefringence, as described in detail in Sec.~\ref{sec:3.2}. The refractive index $n$ of each sample was measured at $25\,{}^{\circ}\mathrm{C}$ using an Abbe refractometer (ER-2S, Matsuyoshi Medical Instruments Co.) because the concentrations of CTAB and NaSal differ among the solutions investigated in this study.

\begin{figure}[b!]
\centering
\includegraphics[width=0.94\columnwidth]{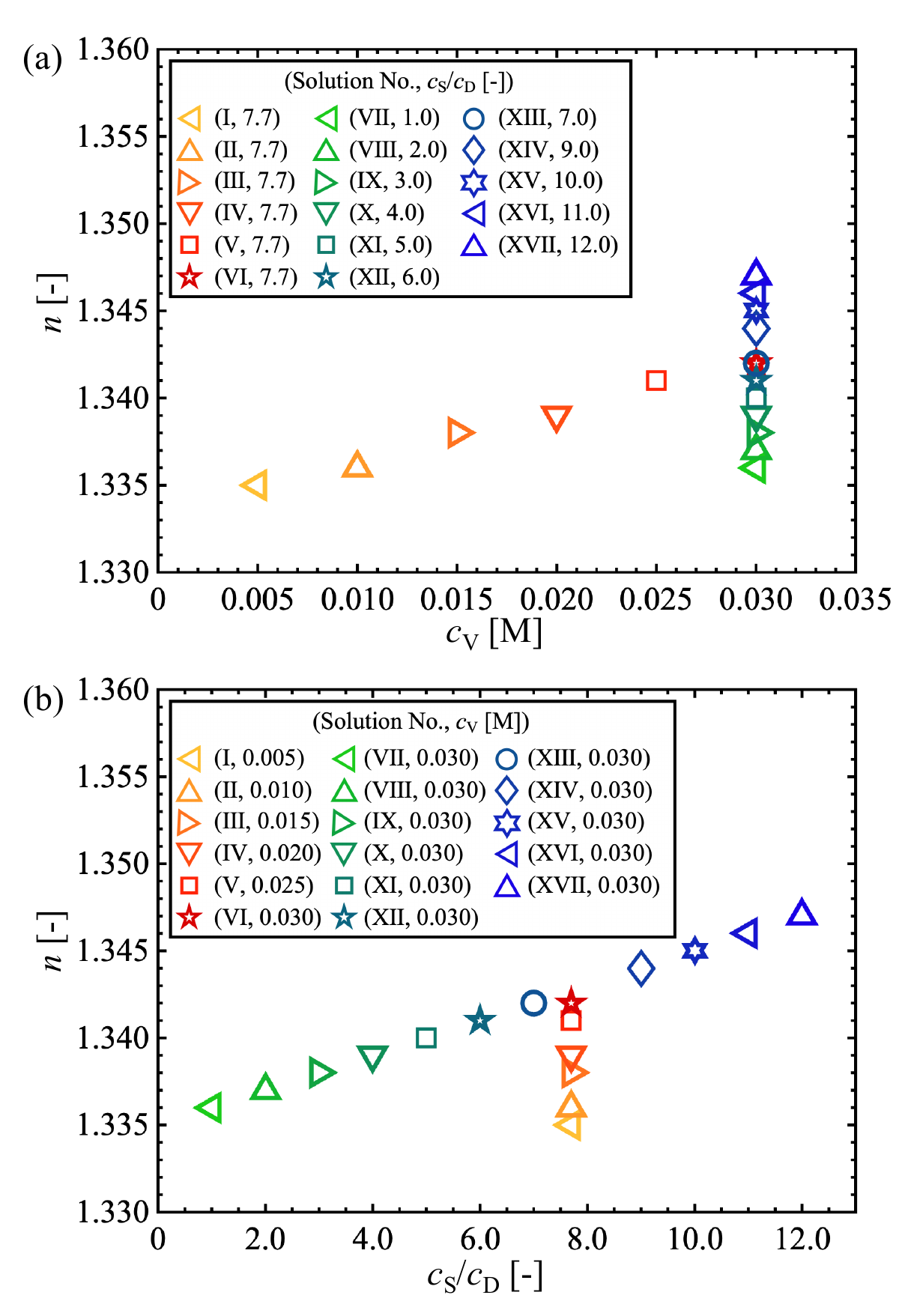}
\caption{
Refractive index $n$ of Solutions I--XVII measured at $25\,{}^{\circ}\mathrm{C}$ plotted against (a) the micellar molar concentration $c_{\mathrm{V}}$ and (b) the CTAB/NaSal molar ratio $c_{\mathrm{S}}/c_{\mathrm{D}}$.}
\label{fig:3}
\end{figure}

Figure~\ref{fig:3} shows the refractive indices of Solutions I--XVII as functions of the micellar molar concentration $c_{\mathrm{V}}$ and the CTAB/NaSal molar ratio $c_{\mathrm{S}}/c_{\mathrm{D}}$. 
As shown in Fig.~\ref{fig:3}(a), the refractive index $n$ increased with increasing $c_{\mathrm{V}}$ for Solutions I--VI, where $c_{\mathrm{S}}/c_{\mathrm{D}}$ was fixed at 7.7. 
In these solutions, both CTAB and NaSal molar concentrations increased with increasing micellar molar concentration $c_{\mathrm{V}}$.
In contrast, Fig.~\ref{fig:3}(b) shows that the refractive index increased with increasing $c_{\mathrm{S}}/c_{\mathrm{D}}$ for Solutions VII--XVII, where $c_{\mathrm{V}}$ was fixed at $0.030\,\mathrm{M}$. In this group, the CTAB concentration remained constant while the NaSal concentration increased, which suggests that the increase in refractive index can be attributed primarily to NaSal. In general, the refractive index of a solution depends on the concentrations and polarizabilities of its constituent components. \cite{fuller1995optical,pearce1931some,vikram1992determination} Thus, the observed trends reflect changes in the optical properties of the solutions associated with variations in CTAB and NaSal concentrations. The effects of refractive index variations on the evaluation of polarizability anisotropy are discussed in Sec.~\ref{sec:3.2}.

\subsection{\label{sec:2.3}Shear viscosity}
The shear viscosity $\eta$ [$\mathrm{Pa\,s}$] of each sample solution was measured using a cone-and-plate rotational rheometer (MCR-302e, Anton Paar GmbH) at $25\,{}^{\circ}\mathrm{C}$. The shear rate $\dot{\gamma}$ [$\mathrm{s}^{-1}$] was varied from $0.1$ to $1000\,\mathrm{s}^{-1}$ and then decreased back to $0.1\,\mathrm{s}^{-1}$. This up-and-down shear rate sweep was repeated three times.

\begin{figure}[b!]
\centering
\includegraphics[width=0.9\columnwidth]{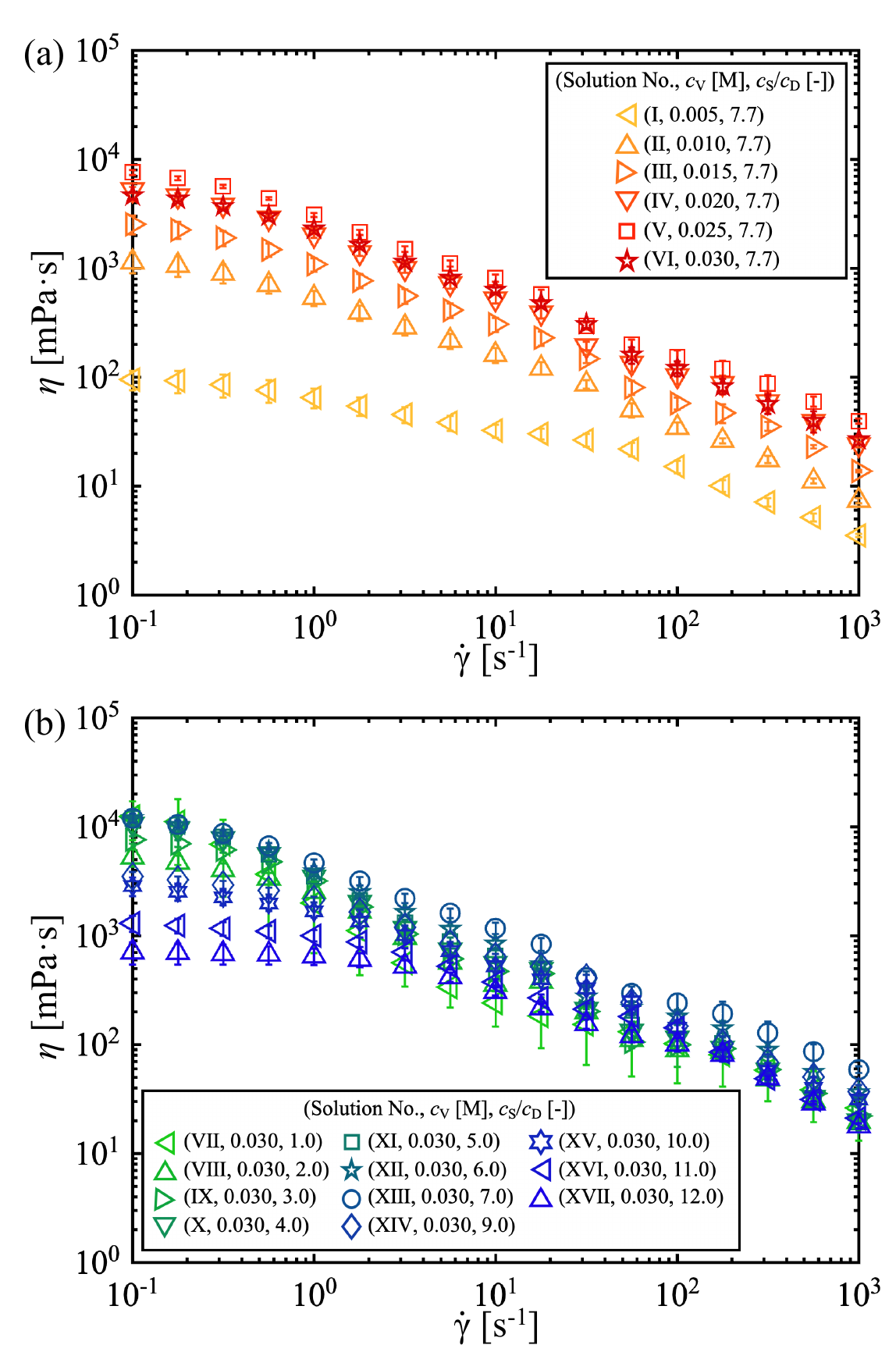}
\caption{
Shear viscosity $\eta$ as a function of shear rate $\dot{\gamma}$ for (a) Solutions I--VI at a constant CTAB/NaSal molar ratio of $c_{\mathrm{S}}/c_{\mathrm{D}} = 7.7$ and (b) Solutions VII--XVII at a constant micellar molar concentration of $c_{\mathrm{V}} = 0.030\,\mathrm{M}$. The symbols and error bars, respectively, represent the mean and standard error obtained from six measurements.}
\label{fig:4}
\end{figure}

Figure~\ref{fig:4}(a) shows the shear viscosities of Solutions I--VI at a fixed CTAB/NaSal molar ratio of $c_{\mathrm{S}}/c_{\mathrm{D}} = 7.7$ with varying micellar molar concentration $c_{\mathrm{V}}$.
Fig.~\ref{fig:4}(b) shows the results for Solutions VII--XVII at a fixed $c_{\mathrm{V}} = 0.030\,\mathrm{M}$, with varying $c_{\mathrm{S}}/c_{\mathrm{D}}$. All solutions exhibited shear-thinning behavior, with viscosity decreasing as the shear rate increased. For Solutions I--VI, the viscosity in the low-shear-rate region generally increased with increasing $c_{\mathrm{V}}$, although Solution VI exhibited a lower viscosity than Solution V. For Solutions VII--XVII, the shear viscosity was relatively high around $c_{\mathrm{S}}/c_{\mathrm{D}} = 5$--$7$ and decreased at $c_{\mathrm{S}}/c_{\mathrm{D}} \geq 9$.

The shear viscosity of CTAB/NaSal aqueous solutions depends not only on the micellar molar concentration but also on the internal structure, including micellar length, entanglement, branching, and cross-linking. \cite{shikata1987micelle,shikata1989micelle,shikata1988micelle,inoue2005nonlinear,lin2001influence,kadoma1997structural,shikata1988nonlinear} Therefore, the nonmonotonic viscosity changes shown in Fig.~\ref{fig:4} may reflect structural differences associated with variations in $c_{\mathrm{V}}$ and $c_{\mathrm{S}}/c_{\mathrm{D}}$. In particular, increasing salt concentration has been reported to induce a transition from entangled wormlike micelles to multiply connected micellar networks. \cite{kadoma1997structural} However, given that the contributions of these structural factors cannot be separated from the present shear viscosity measurements alone, the micellar length $L$ was evaluated separately using dynamic viscoelasticity in Sec.~\ref{sec:4}. In addition, its relationship with the stress-optical response was examined.

\section{\label{sec:3}Evaluation of the stress-optical response under uniaxial extensional flow}
\subsection{\label{sec:3.1}Determination of the stress-optical coefficient using full-field extensional rheo-optical measurements}
\begin{figure}[t!]
\centering
\includegraphics[width=\columnwidth]{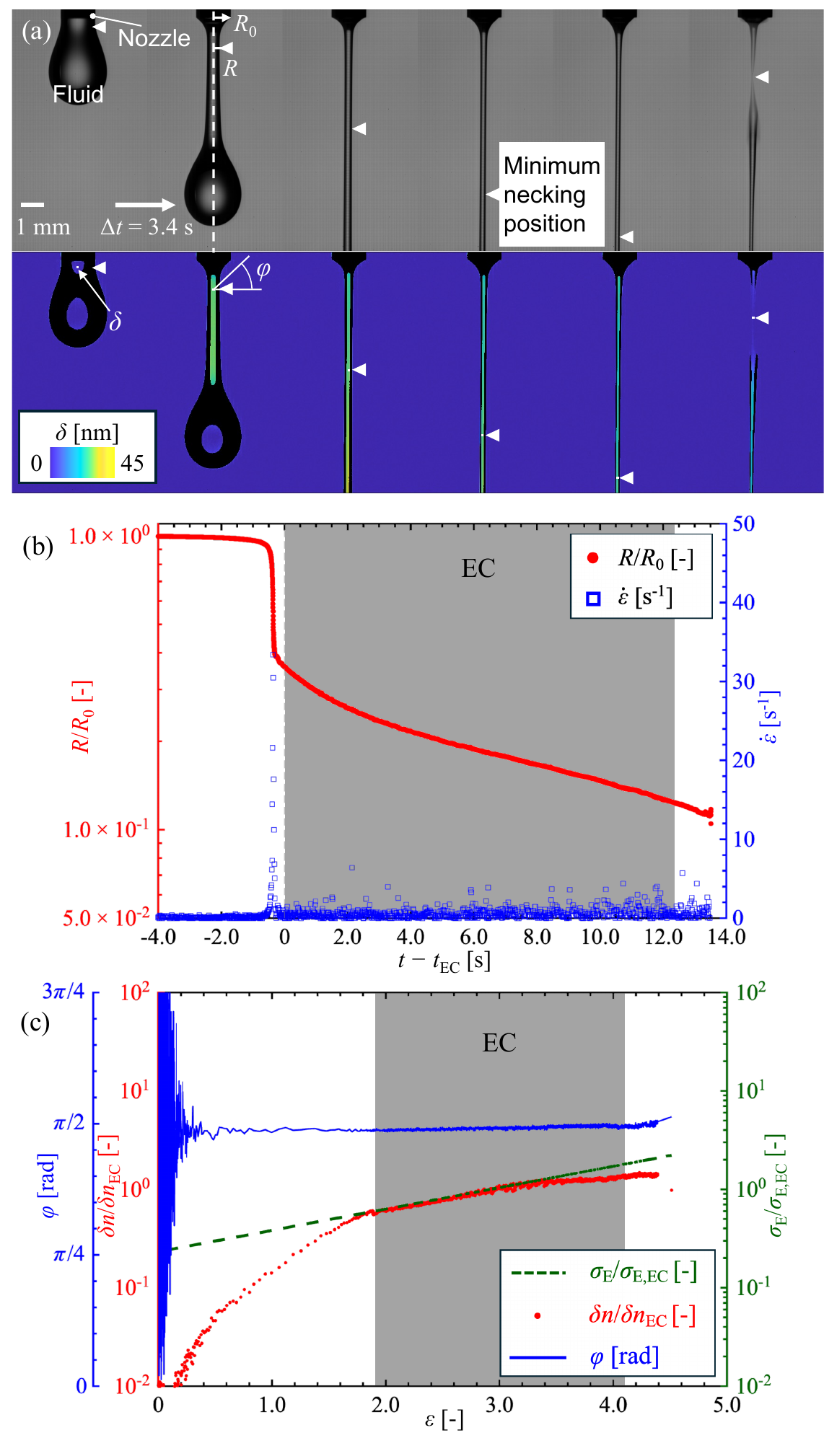}
\caption{
Representative results of the full-field extensional rheo-optical measurements for Solution VI ($c_{\mathrm{V}} = 0.030\,\mathrm{M}$, $c_{\mathrm{S}}/c_{\mathrm{D}} = 7.7$): (a) temporal evolution of light intensity (top) and phase retardation (bottom) images, (b) radius ratio $R/R_0$ and extensional strain rate $\dot{\varepsilon}$ as functions of elapsed time $t - t_{\mathrm{EC}}$ from the onset of the elasto-capillary (EC) regime, and (c) normalized extensional stress $\sigma_{\mathrm{E}}/\sigma_{\mathrm{E,EC}}$, normalized birefringence $\delta n/\delta n_{\mathrm{EC}}$, and orientation angle $\varphi$ as functions of extensional strain $\varepsilon$. The white squares in (a) indicate the measurement area at the filament center.
}
\label{fig:5}
\end{figure}

In this study, the birefringence and extensional stress data of CTAB/NaSal aqueous solutions under uniaxial extensional flow obtained using our previously developed full-field extensional rheo-optical measurement method \cite{muto2025fullfield} were analyzed to evaluate the stress-optical coefficient, polarizability anisotropies, and related parameters. Details of the experimental apparatus, procedures, and analysis methods are provided in our previous study. \cite{muto2025fullfield} Figure~\ref{fig:5} shows representative results for Solution VI under uniaxial extensional flow, which corresponds to a solution composition previously investigated by Shikata et al. \cite{shikata1994rheooptical}

As shown in the light intensity images in the upper part of Fig.~\ref{fig:5}(a), the sample solution dripping from the nozzle tip forms a filament that thins over time. The filament-thinning process was recorded using the high-speed polarization camera (CRYSTA PI-1P, Photonic Lattice, Inc.) to obtain the filament shape, phase retardation $\delta$ [m], and orientation angle $\varphi$ [rad] as shown in the lower part of Fig.~\ref{fig:5}(a). The measured phase retardation $\delta$ is expressed as the integral of the birefringence $\delta n$ [-] along the light-propagation direction $z$:
\begin{equation}
    \delta = \int \delta n \,\mathrm{d}z.
    \label{eq:2}
\end{equation}
Assuming a cylindrical filament with light passing through its center, the optical path length was approximated by the filament diameter $2R$ [m]. The birefringence was also assumed to be uniform along the light propagation direction within the measurement area. A small measurement area was selected at the center of the filament to minimize the lensing effect caused by the curvature of the filament,\cite{muto2025fullfield} as indicated by the white square in Fig.~\ref{fig:5}(a). Using the simultaneously measured filament radius $R$ and phase retardation $\delta$, the birefringence was calculated as
\begin{equation}
    \delta n = \frac{\delta}{2R}.
    \label{eq:3}
\end{equation}

The extensional stress was evaluated from the filament-thinning behavior of the viscoelastic fluid. Before breakup, such filaments undergo the inertio-capillary/visco-capillary (IC/VC), EC, and terminal visco-elasto-capillary (TVEC) regimes. \cite{muto2025fullfield} The extensional stress was evaluated in the EC regime, where elastic stress balances capillary force. In this regime, the filament radius ratio $R(t)/R_0$ [-] decreases exponentially with time $t$ [s] as\cite{omidvar2019detecting,dinic2019macromolecular,dinic2015extensional,mckinley2005visco}
\begin{equation}
    \frac{R(t)}{R_0}
    = \left( \frac{G R_0}{2\Gamma} \right)^{1/3}
    \exp \left[
        -\frac{t - t_{\mathrm{EC}}}{3\lambda_{\mathrm{E}}}
    \right],
    \label{eq:4}
\end{equation}
where $R(t)$ [m] is the minimum filament radius at time $t$, $R_0$ [m] is the nozzle outer radius, $G$ [Pa] is the elastic modulus, $\Gamma$ [$\mathrm{N\,m}^{-1}$] is the surface tension, $\lambda_{\mathrm{E}}$ [s] is the extensional relaxation time, and $t_{\mathrm{EC}}$ [s] is the onset time of the EC regime. The relaxation time $\lambda_{\mathrm{E}}$ was determined by fitting the temporal variation of $R(t)$ in the EC regime with an exponential function.\cite{dinic2019macromolecular}
Assuming incompressible uniaxial extensional flow inside the filament, the extensional strain $\varepsilon$ [-] is obtained from the incompressibility condition as
\begin{equation}
    \varepsilon(t)
    = -2 \ln \left( \frac{R(t)}{R_0} \right).
    \label{eq:5}
\end{equation}
The extensional strain rate $\dot{\varepsilon}$ [s$^{-1}$] is then given by
\begin{equation}
    \dot{\varepsilon}(t)
    = \frac{\mathrm{d}\varepsilon(t)}{\mathrm{d}t}
    = -\frac{2}{R(t)} \frac{\mathrm{d}R(t)}{\mathrm{d}t}.
    \label{eq:6}
\end{equation}
Both $\varepsilon$ and $\dot{\varepsilon}$ were calculated from the measured filament radius $R(t)$.
Figure~\ref{fig:5}(b) shows the temporal variations of $R/R_0$ and $\dot{\varepsilon}$ for Solution VI, with time measured relative to the onset of the EC regime ($t - t_{\mathrm{EC}}$). The IC/VC regime corresponds to $t - t_{\mathrm{EC}} < 0$, whereas the EC and subsequent regimes correspond to $t - t_{\mathrm{EC}} \geq 0$. In the EC regime, $R/R_0$ decreased exponentially, while $\dot{\varepsilon}$ remained nearly constant compared with the other regimes. Under these conditions, the extensional stress $\sigma_{\mathrm{E}}$ [Pa] was evaluated from the balance between elastic stress and capillary force as\cite{anna2001elastocapillary,mckinley2002filament}
\begin{equation}
    \sigma_{\mathrm{E}}(t) = \frac{\Gamma}{R(t)}.
    \label{eq:7}
\end{equation}

In this measurement method, the filament axis coincides with the uniaxial extensional direction, the primary micellar orientation direction, and the optical principal axis. \cite{muto2025fullfield} The axial stress dominates in the EC regime, which allows the principal stress difference to be approximated by the extensional stress $\sigma_{\mathrm{E}}$ as
\begin{equation}
    \sigma_{\parallel} - \sigma_{\perp}
    \simeq \sigma_{\mathrm{E}}.
    \label{eq:8}
\end{equation}
Substituting Eq.~\eqref{eq:8} into the SOR (Eq.~\eqref{eq:1}) gives
\begin{equation}
    \delta n = C \sigma_{\mathrm{E}}.
    \label{eq:9}
\end{equation}
Thus, the stress-optical coefficient $C$ was determined from the slope of the linear relationship between birefringence $\delta n$ and extensional stress $\sigma_{\mathrm{E}}$, measured simultaneously in the EC regime.\cite{muto2025fullfield}

Figure~\ref{fig:5}(c) shows the normalized extensional stress $\sigma_{\mathrm{E}}/\sigma_{\mathrm{E,EC}}$ [-], normalized birefringence $\delta n/\delta n_{\mathrm{EC}}$ [-], and orientation angle $\varphi$ [rad] as functions of extensional strain $\varepsilon$. Here, $\sigma_{\mathrm{E,EC}}$ [Pa] and $\delta n_{\mathrm{EC}}$ [-] respectively denote the extensional stress and birefringence at the onset of the EC regime. If the stress-optical coefficient $C$ [Pa$^{-1}$] remains constant throughout the EC regime, Eq.~\eqref{eq:9} implies that the normalized extensional stress and birefringence should coincide: ($\sigma_{\mathrm{E}}/\sigma_{\mathrm{E,EC}} = \delta n/\delta n_{\mathrm{EC}}$).
The normalized birefringence was initially small and increased as the filament thinned, which indicates the development of optical anisotropy associated with micellar orientation. In the IC/VC regime, the normalized extensional stress and birefringence exhibited different trends but approached each other near the onset of the EC regime and approximately coincided within this regime. The orientation angle $\varphi$ measured from the horizontal direction ($0\ \mathrm{rad}$) approached $\pi/2\ \mathrm{rad}$, corresponding to the extensional direction, before the onset of the EC regime and remained nearly constant thereafter. These results indicate that the wormlike micelles were predominantly aligned along the extensional direction and that the extensional stress and birefringence varied approximately proportionally in the EC regime. \cite{muto2025fullfield} Thus, the SOR was considered valid in this regime, which allowed the stress-optical coefficient $C$ to be evaluated.

These results demonstrate that the stress-optical coefficient $C$ can be evaluated from the relationship between birefringence $\delta n$ and extensional stress $\sigma_{\mathrm{E}}$ in the EC regime. However, birefringence depends on the orientation of the wormlike micelles under flow and their intrinsic optical anisotropy. To account for this effect, birefringence $\delta n$ can be expressed in terms of the intrinsic birefringence $\delta n^{0}$ [-] and orientation parameter $f$ [-] as \cite{humbert1998stress,humbert1998flow,wheeler1996structure,shikata1994rheooptical,decruppe2003flow,sridhar2000birefringence}
\begin{equation}
    \delta n = f \delta n^{0}.
    \label{eq:10}
\end{equation}
Here, $\delta n^{0}$ represents the birefringence under perfect micellar alignment along the extensional direction.\cite{muto2025fullfield} In the present analysis, the orientation parameter $f$ was estimated from the measured optical principal-axis orientation $\varphi$ as\cite{okada2016reliability}
\begin{equation}
    f = \frac{3\cos^{2}\left(\pi/2 - \varphi\right) - 1}{2},
    \label{eq:11}
\end{equation}
where $\varphi$ is measured from the horizontal direction, with $\pi/2\ \mathrm{rad}$ corresponding to the extensional direction. As shown in Fig.~\ref{fig:5}(c), $\varphi$ remained nearly constant around $\pi/2\ \mathrm{rad}$ in the EC regime. Similar behavior was observed for all sample solutions, \cite{muto2025fullfield} which indicates that the wormlike micelles were statistically preferentially oriented along the extensional direction. Although Eq.~\eqref{eq:11} yields $f \simeq 1$ under these conditions, this does not necessarily imply perfect micellar alignment because individual micelles may exhibit a distribution of orientation angles. Therefore, assuming $f \simeq 1$ in the EC regime, the birefringence can be approximated as $\delta n \simeq \delta n^{0}$, and Eq.~\eqref{eq:9} becomes
\begin{equation}
    \delta n^{0} \simeq C \sigma_{\mathrm{E}}.
    \label{eq:12}
\end{equation}

\begin{figure}[t]
\centering
\includegraphics[width=0.92\columnwidth]{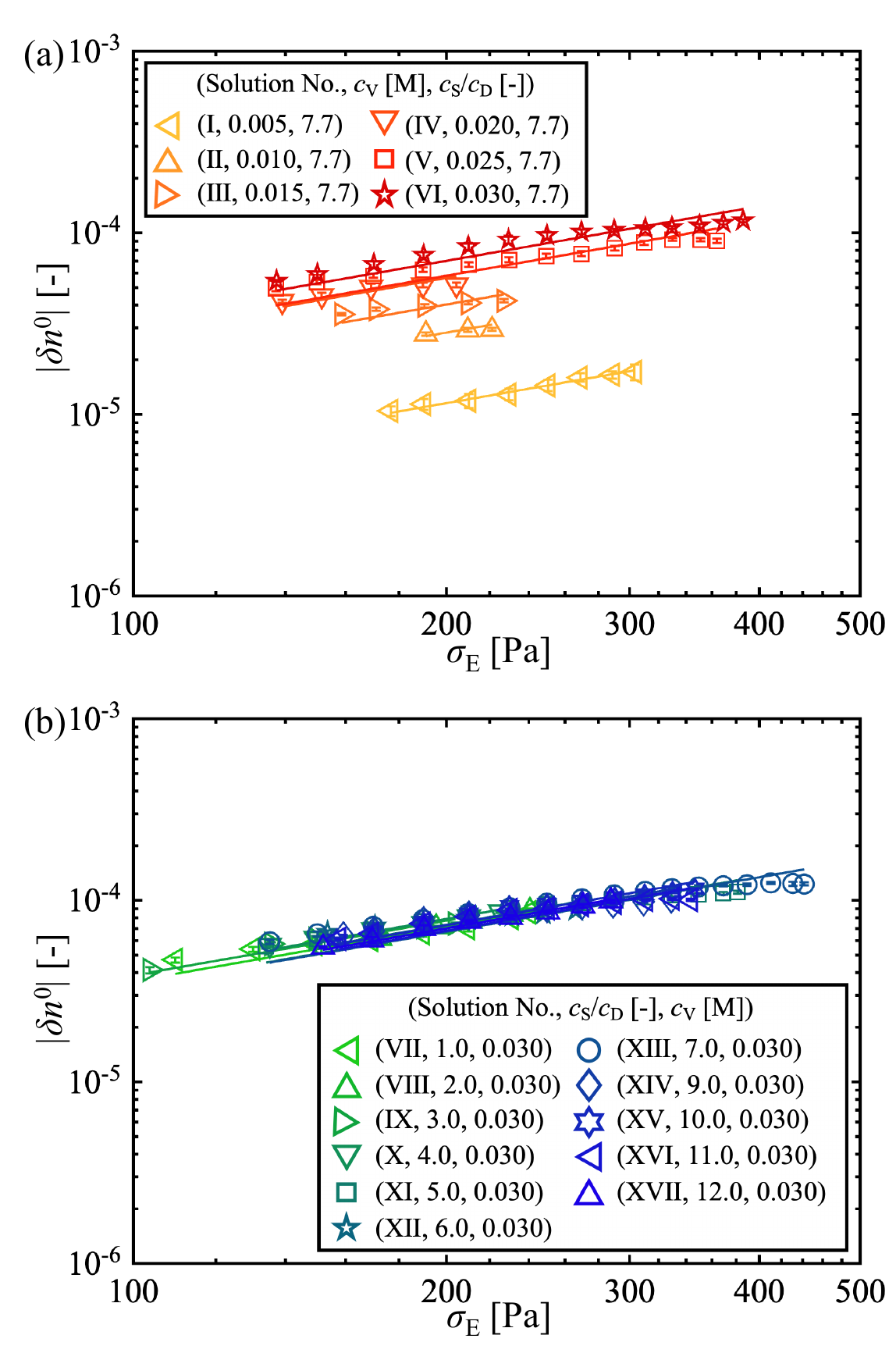}
\caption{
Relationship between intrinsic birefringence $|\delta n^{0}|$ and extensional stress $\sigma_{\mathrm{E}}$ in the EC regime for (a) Solutions I--VI at $c_{\mathrm{S}}/c_{\mathrm{D}} = 7.7$ and (b) Solutions VII--XVII at $c_{\mathrm{V}} = 0.030\,\mathrm{M}$.}
\label{fig:6}
\end{figure}

Figure~\ref{fig:6} shows the relationship between the extensional stress $\sigma_{\mathrm{E}}$ and intrinsic birefringence $|\delta n^{0}|$ in the EC regime. 
Figure~\ref{fig:6}(a) shows the results for Solutions I--VI, for which the CTAB/NaSal molar ratio was fixed at $c_{\mathrm{S}}/c_{\mathrm{D}} = 7.7$ and the micellar molar concentration $c_{\mathrm{V}}$ was varied. Figure~\ref{fig:6}(b) shows the results for Solutions VII--XVII, for which $c_{\mathrm{V}}$ was fixed at $0.030\,\mathrm{M}$ and the CTAB/NaSal molar ratio $c_{\mathrm{S}}/c_{\mathrm{D}}$ was varied. For all sample solutions, $|\delta n^{0}|$ increased approximately linearly with increasing $\sigma_{\mathrm{E}}$. These results confirm that the SOR given by Eq.~\eqref{eq:12} holds in the EC regime for all the sample solutions used in this study. Accordingly, the stress-optical coefficient $C$ was calculated for each sample solution from the slope of the linear relationship between $\delta n^{0}$ and $\sigma_{\mathrm{E}}$. The obtained values of $C$ are discussed in Sec.~\ref{sec:3.2}.

\subsection{\label{sec:3.2}Calculation of polarizability anisotropy}
In the previous section, the stress-optical coefficient $C$ was evaluated from the relationship between intrinsic birefringence $\delta n^{0}$ and extensional stress $\sigma_{\mathrm{E}}$. Here, the polarizability anisotropies of equivalent Kuhn segments and micellar segments are evaluated using relationships derived for Gaussian polymer chain networks and applied to CTAB/NaSal wormlike micelles by Shikata et al., \cite{shikata1994rheooptical} to examine the stress-optical response in terms of microscopic optical properties.

The polarizability anisotropy $\Delta\alpha$ [\AA$^{3}$] of an equivalent Kuhn segment was calculated from the stress-optical coefficient $C$ as \cite{shikata1994rheooptical}
\begin{equation}
    \Delta\alpha
    = \frac{45k_{\mathrm{B}}\bar{n}T}
    {2\pi\left(\bar{n}^{2}+2\right)^{2}} C.
    \label{eq:13}
\end{equation}
Here, $\bar{n}$ [-] is the mean refractive index of the solution, $k_{\mathrm{B}}$ [$\mathrm{J\,K}^{-1}$] is the Boltzmann constant, and $T$ [K] is the absolute temperature. The quantity $\Delta\alpha$ represents the difference between the longitudinal and transverse polarizabilities of an equivalent Kuhn segment. In contrast, the polarizability anisotropy $\Delta\alpha^{0}$ [\AA$^{3}$] of a micellar segment was calculated from the intrinsic birefringence $\delta n^{0}$ and micellar molar concentration $c_{\mathrm{V}}$ as \cite{shikata1994rheooptical}
\begin{equation}
    \Delta\alpha^{0}
    = \frac{9m\bar{n}}
    {2\pi\left(\bar{n}^{2}+2\right)^{2}N_{\mathrm{A}}}
    \frac{\delta n^{0}}{c_{\mathrm{V}}}.
    \label{eq:14}
\end{equation}
Here, $N_{\mathrm{A}}$ [$\mathrm{mol}^{-1}$] is the Avogadro constant, and $m$ is the number of micellar units ($\mathrm{CTA}^{+}\mathrm{Sal}^{-}$ complexes) per micellar segment. Thus, $\Delta\alpha$ characterizes the optical anisotropy of an equivalent Kuhn segment accounting for micellar flexibility, whereas $\Delta\alpha^{0}$ represents that of an individual micellar segment.

The polarizability anisotropies of equivalent Kuhn segments and micellar segments were calculated using Eqs.~\eqref{eq:13} and \eqref{eq:14} with $T = 298\,\mathrm{K}$, $k_{\mathrm{B}} = 1.38 \times 10^{-23}\,\mathrm{J\,K}^{-1}$, and $N_{\mathrm{A}} = 6.02 \times 10^{23}\,\mathrm{mol}^{-1}$. The effects of variations in the refractive index among the samples were also examined. As shown in Sec.~\ref{sec:2.2}, the measured refractive indices ranged from $1.335$ to $1.347$, with a mean of 1.340 across all 17 samples. The refractive-index-dependent term common to both equations is given by\cite{shikata1994rheooptical}
\begin{equation}
    F(\bar{n}) = \frac{\bar{n}}{\left(\bar{n}^{2}+2\right)^{2}}.
    \label{eq:15}
\end{equation}
The value of $F(\bar{n})$ at $\bar{n} = 1.340$ was 0.09301, compared with 0.09332 and 0.09258 at the minimum and maximum measured refractive indices ($n = 1.335$ and $1.347$), respectively. Using $\bar{n} = 1.340$ instead of the individual measured refractive indices resulted in a maximum relative error of 0.47\% in the calculated polarizability anisotropies. Thus, variations in the refractive index among the samples had a negligible effect on $\Delta\alpha$ and $\Delta\alpha^{0}$, and $\bar{n} = 1.340$ was used for all samples in the subsequent analyses.

Table~\ref{tab:2} lists the calculated intrinsic birefringence $\delta n^{0}$, stress-optical coefficient $C$, and polarizability anisotropies $\Delta\alpha$ and $\Delta\alpha^{0}$, including their signs. 
\begin{table}[bt!]
    \centering
    \caption{
    Intrinsic birefringence $\delta n^{0}$, stress-optical coefficient $C$, and polarizability anisotropies $\Delta\alpha$ and $\Delta\alpha^{0}$ of equivalent Kuhn and micellar segments, respectively, for Solutions I--XVII. The quantities $\delta n^{0}$, $\Delta\alpha$, and $\Delta\alpha^{0}$ were evaluated using Eqs.~\eqref{eq:12}--\eqref{eq:14}, respectively. Values are reported as estimates $\pm$ associated uncertainties.}
    \label{tab:2}
    \setlength{\tabcolsep}{3pt}
    \renewcommand{\arraystretch}{1.35}
    \begin{tabular*}{\columnwidth}{@{\extracolsep{\fill}}lcccc@{}}
    \hline\hline
    Solution
    & $\delta n^{0} \times 10^{5}$
    & $C \times 10^{7}$
    & $\Delta\alpha$
    & $\Delta\alpha^{0}$ \\
    No.
    & [-]
    & [$\mathrm{Pa}^{-1}$]
    & [\AA$^{3}$]
    & [\AA$^{3}$] \\
    \hline
    I    & $-1.37 \pm 0.22$ & $-0.58 \pm 0.03$ & $-159 \pm 11$   & $-11.0 \pm 1.8$ \\
    II   & $-2.86 \pm 0.10$ & $-1.41 \pm 0.05$ & $-387 \pm 14$   & $-11.4 \pm 0.4$ \\
    III  & $-4.01 \pm 0.19$ & $-2.04 \pm 0.13$ & $-560 \pm 36$   & $-10.7 \pm 0.5$ \\
    IV   & $-4.91 \pm 0.35$ & $-2.98 \pm 0.17$ & $-784 \pm 51$   & $-9.8 \pm 0.7$  \\
    V    & $-7.37 \pm 1.26$ & $-3.04 \pm 0.30$ & $-833 \pm 81$   & $-11.7 \pm 2.0$ \\
    VI   & $-9.38 \pm 1.66$ & $-3.67 \pm 0.35$ & $-1000 \pm 97$  & $-12.4 \pm 2.2$ \\
    VII  & $-5.90 \pm 1.05$ & $-3.85 \pm 0.36$ & $-1060 \pm 99$  & $-7.9 \pm 1.4$  \\
    VIII & $-7.96 \pm 0.53$ & $-3.63 \pm 0.06$ & $-996 \pm 16$   & $-10.6 \pm 0.7$ \\
    IX   & $-6.66 \pm 0.67$ & $-3.89 \pm 0.21$ & $-1070 \pm 56$  & $-8.9 \pm 0.9$  \\
    X    & $-7.16 \pm 0.98$ & $-3.99 \pm 0.16$ & $-1090 \pm 43$  & $-9.5 \pm 1.3$  \\
    XI   & $-8.50 \pm 1.73$ & $-3.79 \pm 0.38$ & $-1040 \pm 104$ & $-11.3 \pm 2.3$ \\
    XII  & $-7.92 \pm 0.60$ & $-3.78 \pm 0.37$ & $-1040 \pm 101$ & $-10.5 \pm 0.8$ \\
    XIII & $-9.58 \pm 1.96$ & $-3.80 \pm 0.42$ & $-1040 \pm 114$ & $-12.7 \pm 2.6$ \\
    XIV  & $-8.93 \pm 0.98$ & $-3.60 \pm 0.35$ & $-983 \pm 95$   & $-11.8 \pm 1.3$ \\
    XV   & $-9.59 \pm 1.59$ & $-3.73 \pm 0.27$ & $-1020 \pm 75$  & $-12.7 \pm 2.1$ \\
    XVI  & $-8.86 \pm 1.21$ & $-3.53 \pm 0.32$ & $-967 \pm 88$   & $-11.7 \pm 1.6$ \\
    XVII & $-7.88 \pm 1.44$ & $-3.53 \pm 0.11$ & $-963 \pm 29$   & $-10.4 \pm 1.9$ \\
    \hline\hline
    \end{tabular*}
\end{table}
The absolute values $|\delta n^{0}|$, $|C|$, $|\Delta\alpha|$, and $|\Delta\alpha^{0}|$ are used in the subsequent analyses of their dependences on the CTAB/NaSal molar ratio and micellar molar concentration. The uncertainties in $\Delta\alpha$ and $\Delta\alpha^{0}$ were determined by propagating those of $C$ and $\delta n^{0}$ through Eqs.~\eqref{eq:13} and \eqref{eq:14}, respectively.

The magnitude of the polarizability anisotropy of micellar segments $|\Delta\alpha^{0}|$ ranged from $7.9$ to $12.7$~\AA$^{3}$ in this study, comparable to the value of approximately $14$~\AA$^{3}$ reported by Shikata et al. \cite{shikata1994rheooptical} for CTAB/NaSal wormlike micelles with $m = 18$. Decruppe and Ponton \cite{decruppe2003flow} also reported that the polarizability anisotropy of micellar segments in CTAC/NaSal wormlike micelles showed little dependence on salt concentration. For comparison, reported polarizability anisotropies include approximately $2.6$~\AA$^{3}$ for carbon dioxide, \cite{gough1983laser} $6.0$, $9.5$, and $20$~\AA$^{3}$ for benzene, naphthalene, and anthracene, respectively, \cite{ren2011polarizable} and $65$--$81$~\AA$^{3}$ for bent-core nematic liquid-crystal molecules. \cite{upadhyaya2013a} Although direct quantitative comparisons are limited by differences in the molecular structures and definitions of polarizability anisotropy, these values provide a reference to assess the variation in $|\Delta\alpha^{0}|$ observed in this study. The variation from $7.9$ to $12.7$~\AA$^{3}$ was relatively small in this broader context, with no clear systematic dependence on the CTAB/NaSal molar ratio or micellar molar concentration. Therefore, within the experimental accuracy of this study, $|\Delta\alpha^{0}|$ was regarded as approximately constant across the sample conditions.

\subsection{\label{sec:3.3}Dependence of the stress-optical response on molar ratio and micellar molar concentration}
This section examines the dependences of intrinsic birefringence, stress-optical coefficient, and polarizability anisotropies on micellar molar concentration and CTAB/NaSal molar ratio.

\begin{figure}[t!]
\centering
\includegraphics[width=0.94\columnwidth]{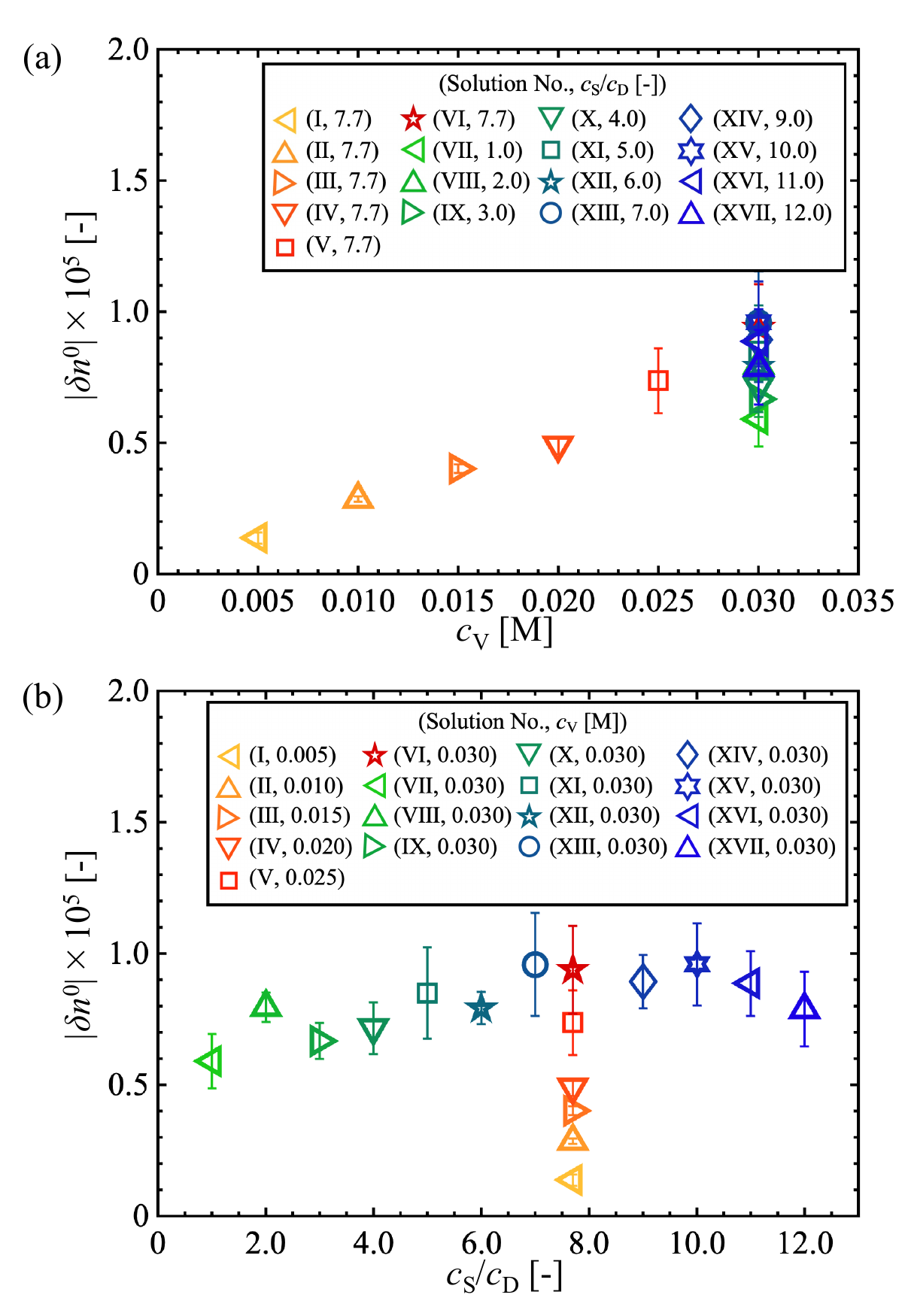}
\caption{
Dependence of the magnitude of the intrinsic birefringence $|\delta n^{0}|$ on (a) micellar molar concentration $c_{\mathrm{V}}$ and (b) CTAB/NaSal molar ratio $c_{\mathrm{S}}/c_{\mathrm{D}}$ for Solutions I--XVII. The same experimental data are plotted in both panels using different horizontal axes.}
\label{fig:7}
\end{figure}

Figure~\ref{fig:7} shows the dependence of the magnitude of the intrinsic birefringence $|\delta n^{0}|$ on (a) the micellar molar concentration $c_{\mathrm{V}}$ and (b) the CTAB/NaSal molar ratio $c_{\mathrm{S}}/c_{\mathrm{D}}$.
As shown in Fig.~\ref{fig:7}(a), $|\delta n^{0}|$ increased with increasing $c_{\mathrm{V}}$ for Solutions I--VI at a fixed $c_{\mathrm{S}}/c_{\mathrm{D}} = 7.7$. In contrast, for Solutions VII--XVII, at a fixed $c_{\mathrm{V}} = 0.030\,\mathrm{M}$, $|\delta n^{0}|$ varied among the samples but showed no clear systematic dependence on $c_{\mathrm{S}}/c_{\mathrm{D}}$, as shown in Fig.~\ref{fig:7}(b). However, according to Eq.~\eqref{eq:12}, $\delta n^{0}$ is proportional to the extensional stress $\sigma_{\mathrm{E}}$, which depends on the viscosity, relaxation properties, and filament-thinning behavior of the solution. Thus, differences in $|\delta n^{0}|$ among the samples reflect variations in both optical properties and extensional stress. Consequently, the effects of $c_{\mathrm{V}}$ and $c_{\mathrm{S}}/c_{\mathrm{D}}$ on the stress-optical response cannot be evaluated solely from $|\delta n^{0}|$.

\begin{figure}[tb!]
\centering
\includegraphics[width=0.96\columnwidth]{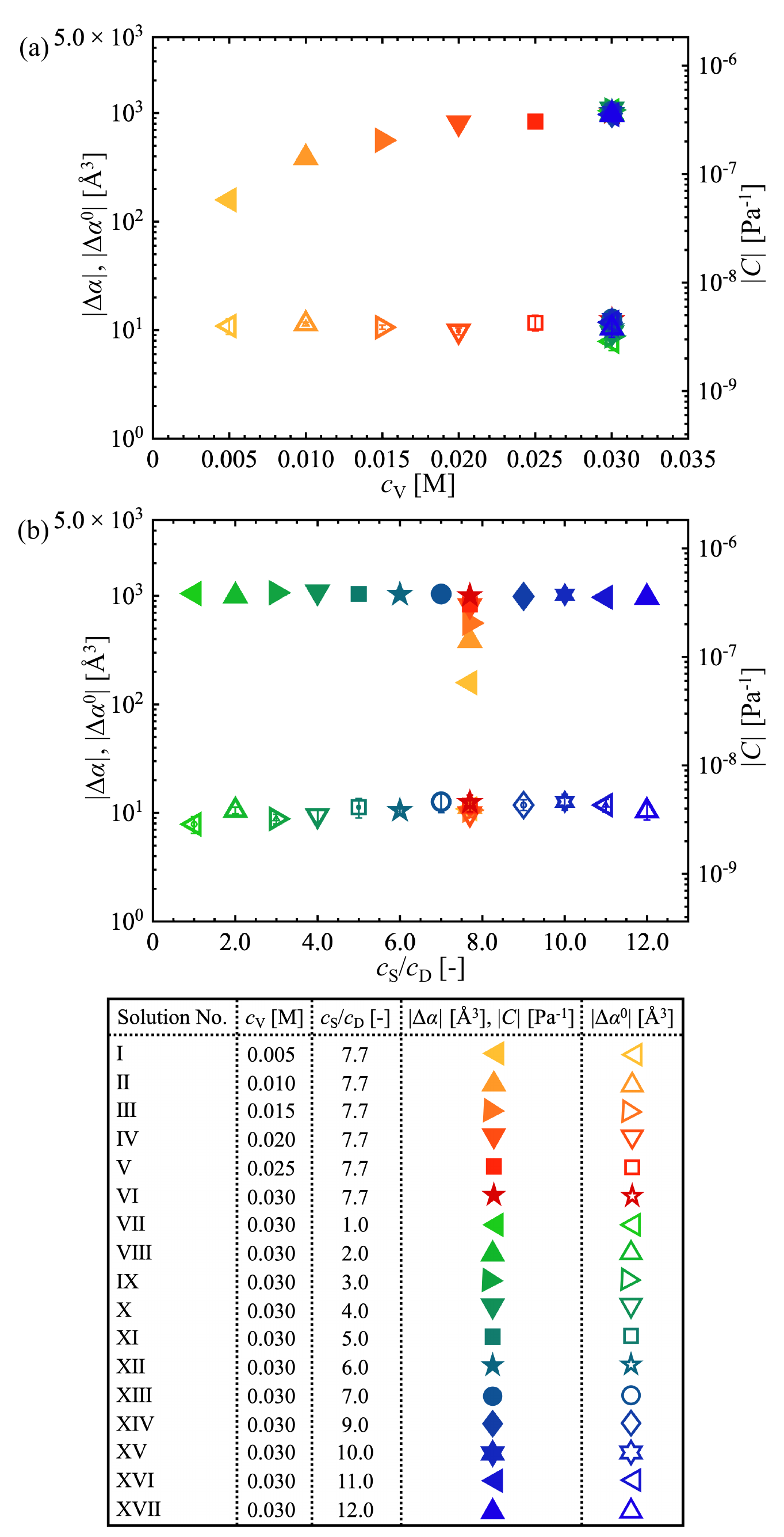}
\caption{
Dependence of the magnitudes of the polarizability anisotropies $|\Delta\alpha|$ and $|\Delta\alpha^{0}|$ and the stress-optical coefficient $|C|$ on (a) micellar molar concentration $c_{\mathrm{V}}$ and (b) CTAB/NaSal molar ratio $c_{\mathrm{S}}/c_{\mathrm{D}}$ for Solutions I--XVII. The same experimental data are plotted in both panels using different horizontal axes. The polarizability anisotropies are plotted on the left vertical axis and $|C|$ on the right vertical axis. The two vertical-axis scales are related by the proportionality between $|\Delta\alpha|$ and $|C|$ in Eq.~\eqref{eq:13}, allowing both quantities to share the same data points. The legend for both panels is placed below panel (b).}
\label{fig:8}
\end{figure}

Therefore, attention was focused on the stress-optical coefficient $C$, which relates intrinsic birefringence to extensional stress, and the polarizability anisotropies $\Delta\alpha$ and $\Delta\alpha^{0}$ of equivalent Kuhn segments and micellar segments, respectively, which characterize the optical anisotropy at microscopic structural scales. Figure~\ref{fig:8} shows the magnitudes $|C|$, $|\Delta\alpha|$, and $|\Delta\alpha^{0}|$.
For Solutions I--VI, at a fixed $c_{\mathrm{S}}/c_{\mathrm{D}} = 7.7$, Fig.~\ref{fig:8}(a) shows that $|C|$ and $|\Delta\alpha|$ increased with increasing $c_{\mathrm{V}}$. The two quantities exhibited corresponding changes because they are proportional according to Eq.~\eqref{eq:13}, with the same refractive index used for all samples. In contrast, $|\Delta\alpha^{0}|$ showed no clear systematic dependence on $c_{\mathrm{V}}$.
For Solutions VII--XVII, at a fixed $c_{\mathrm{V}} = 0.030\,\mathrm{M}$, Fig.~\ref{fig:8}(b) shows that $|C|$ and $|\Delta\alpha|$ varied among the samples but exhibited no clear systematic dependence on $c_{\mathrm{S}}/c_{\mathrm{D}}$. Similarly, $|\Delta\alpha^{0}|$ showed no clear dependence on this molar ratio.
These results indicate that the stress-optical response depends strongly on $c_{\mathrm{V}}$ at a fixed $c_{\mathrm{S}}/c_{\mathrm{D}}$ within the investigated range, whereas no clear systematic dependence on $c_{\mathrm{S}}/c_{\mathrm{D}}$ was observed at a fixed $c_{\mathrm{V}}$.

An important finding is that $|\Delta\alpha^{0}|$ remained approximately constant across Solutions I--XVII despite differences in solution composition. As shown in Eq.~\eqref{eq:14}, $|\Delta\alpha^{0}|$ is proportional to the intrinsic birefringence $|\delta n^{0}|$ normalized by the micellar molar concentration $c_{\mathrm{V}}$. Shikata et al. \cite{shikata1994rheooptical} suggested that the optical anisotropy of CTAB/NaSal wormlike micelles mainly originates from the polarizability anisotropy of $\mathrm{CTA}^{+}$ ions constituting the micelles, with a relatively small contribution from $\mathrm{Sal}^{-}$. The approximately constant $|\Delta\alpha^{0}|$ suggests that the local optical anisotropy of individual micellar segments is relatively insensitive to changes in $c_{\mathrm{V}}$ and $c_{\mathrm{S}}/c_{\mathrm{D}}$.

To examine whether the present findings extend to other micellar morphologies, additional measurements were performed for solutions expected to contain spherical micelles (Fig.~\ref{fig:1}), although the detailed results are not presented here. Specifically, five solution compositions were examined at a fixed NaSal molar concentration of $c_{\mathrm{S}} = 0.030\,\mathrm{M}$, with CTAB/NaSal molar ratios $c_{\mathrm{S}}/c_{\mathrm{D}}$ of 0.14, 0.16, 0.18, 0.22, and 0.24. For all five conditions, no clear proportional relationship between birefringence and extensional stress was observed in the intervals identified as the EC regime. Furthermore, the orientation angle did not remain constant near $\pi/2\ \mathrm{rad}$, corresponding to the extensional direction. Consequently, the stress-optical coefficient could not be reliably determined for these solutions using the present evaluation method. Therefore, the findings obtained for the investigated solutions containing wormlike and potentially networked micelles should not necessarily be generalized to all micellar morphologies, including spherical micelles.

The number density of micellar segments per unit volume is given by $10^{3}c_{\mathrm{V}}N_{\mathrm{A}}/m$, which is proportional to $c_{\mathrm{V}}$ when $m$ is assumed to be constant. Furthermore, Eq.~\eqref{eq:14} indicates that the intrinsic birefringence $\delta n^{0}$ is proportional to the product of $c_{\mathrm{V}}$ and the polarizability anisotropy $\Delta\alpha^{0}$ of individual micellar segments. Given that $|\Delta\alpha^{0}|$ showed no clear systematic dependence on the solution composition, the increase in $|\delta n^{0}|$ with $c_{\mathrm{V}}$ is consistent with the increase in micellar segment number density assuming comparable micellar orientation among the samples. This interpretation is also consistent with the concept of orientation birefringence arising from the alignment of optically anisotropic molecules or structural units under flow. \cite{shikata1994rheooptical,fuller1995optical}
In contrast, $|C|$ and $|\Delta\alpha|$ also increased with $c_{\mathrm{V}}$, but their concentration dependence cannot necessarily be explained solely by the increase in micellar segment number density. Since these quantities characterize the stress-optical response and the optical anisotropy of equivalent Kuhn segments, respectively, their variations may also reflect changes in the overall micellar structure, particularly the micellar length $L$. Therefore, examining the relationship between $L$ and the stress-optical parameters is important to understand the physical origin of their concentration dependence.

\section{\label{sec:4}Evaluation of micellar length and its dependence on solution composition}
\subsection{\label{sec:4.1}Method for calculating micellar length and its validation}
The micellar length $L$ representing the average contour length of wormlike micelles illustrated in Fig.~\ref{fig:2} is an important structural parameter that influences the relaxation time and viscosity of the solution through wormlike micellar entanglement and scission--recombination processes. \cite{liberatore2009microstructure,dreiss2007wormlike,walker2001rheology,lequeux1996structure,cates1990statics,rehage1991viscoelastic,turner1992linear,granek1992stress,turner1991linear,cates1987reptation} However, direct determination of micellar length in highly concentrated wormlike micellar solutions using scattering and other techniques is particularly difficult. Therefore, methods to estimate the micellar length from dynamic viscoelastic data have been proposed. \cite{granek1992stress,turner1991linear,cates1987reptation,tan2022quantitative,tan2021determining} Tan et al. \cite{tan2022quantitative,tan2021determining} proposed a modified scaling law based on a pointer algorithm accounting for reptation, Rouse motion, bending motion, and micellar scission--recombination processes. This approach estimates the micellar length from the frequency dependences of the storage modulus $G'(\omega)$ [Pa] and loss modulus $G''(\omega)$ [Pa]. The estimated lengths were reported to be greater than those obtained using the conventional Cates scaling law proposed by Granek and Cates, \cite{granek1992stress} which provides physically reasonable estimates for highly viscous wormlike micellar solutions. \cite{tan2022quantitative,tan2021determining} In this study, the micellar length $L$ of each sample solution was evaluated using this modified scaling law.

The entanglement length $l_{\mathrm{e}}$ [m], which characterizes the entanglement state of wormlike micelles, must first be determined to estimate the micellar length $L$ from the scaling law. The entanglement length $l_{\mathrm{e}}$ represents the contour distance between neighboring entanglement points along a wormlike micelle.\cite{turner1993dynamics, larson1999structure} In this study, $l_{\mathrm{e}}$ was evaluated using the plateau modulus $G_{0}$ [Pa] and the polarizability anisotropies obtained in Sec.~\ref{sec:3.2}. First, the correlation length $\xi$ [m] representing a characteristic distance between micelles was calculated from $G_{0}$ as
\begin{equation}
    \xi = \left(\frac{k_{\mathrm{B}}T}{G_0}\right)^{1/3}.
    \label{eq:16}
\end{equation}
Next, the persistence length $l_{\mathrm{p}}$ [m], which characterizes micellar flexibility, was evaluated using the Kuhn length $b_{\mathrm{K}} = 2l_{\mathrm{p}}$ shown in Fig.~\ref{fig:2}. \cite{yamakawa1971modern, nakayayaegashi2008linear, larson1999structure, degennes1979scaling} Assuming that an equivalent Kuhn segment consists of $b_{\mathrm{K}}/w$ micellar segments, their polarizability anisotropies are related by
\begin{equation}
    |\Delta\alpha|
    = \frac{b_{\mathrm{K}}}{w}|\Delta\alpha^{0}|
    = \frac{2l_{\mathrm{p}}}{w}|\Delta\alpha^{0}|.
    \label{eq:17}
\end{equation}
Here, $w$ [m] is the axial length of a micellar segment, which was set to $0.85\,\mathrm{nm}$ following Shikata et al. \cite{shikata1994rheooptical} Substituting Eqs.~\eqref{eq:13} and \eqref{eq:14} into Eq.~\eqref{eq:17} gives
\begin{equation}
    l_{\mathrm{p}}
    = \frac{|\Delta\alpha|}{2|\Delta\alpha^{0}|}w
    = \frac{5k_{\mathrm{B}}N_{\mathrm{A}}Tw}{2m}
      \frac{|C|c_{\mathrm{V}}}{|\delta n^{0}|}.
    \label{eq:18}
\end{equation}
Using the SOR ($\delta n^{0} \simeq C\sigma_{\mathrm{E}}$) given by Eq.~\eqref{eq:12}, the ratio $|C|/|\delta n^{0}|$ can be expressed as $1/\sigma_{\mathrm{E}}$, which yields $l_{\mathrm{p}} \propto c_{\mathrm{V}}/\sigma_{\mathrm{E}}$. Finally, the entanglement length $l_{\mathrm{e}}$ was calculated from $\xi$ (Eq.~\eqref{eq:16}) and $l_{\mathrm{p}}$ (Eq.~\eqref{eq:18}) as
\begin{equation}
    l_{\mathrm{e}}
    = \frac{\xi^{5/3}}{l_{\mathrm{p}}^{2/3}}
    \simeq \frac{1}{\left(G_0^{5}k_{\mathrm{B}}T\right)^{1/9}}
      \left(
        \frac{2m}{5N_{\mathrm{A}}w}
        \frac{\sigma_{\mathrm{E}}}{c_{\mathrm{V}}}
      \right)^{2/3}.
    \label{eq:19}
\end{equation}

The entanglement length $l_{\mathrm{e}}$ evaluated above was then used to estimate the micellar length $L$ through a scaling law relating $L/l_{\mathrm{e}}$ to the dynamic viscoelastic properties. The conventional scaling law proposed by Granek and Cates \cite{granek1992stress} is expressed as
\begin{equation}
    \frac{G'_{\mathrm{min}}}{G''_{\mathrm{min}}}
    = \left(\frac{L}{l_{\mathrm{e}}}\right)^{0.8},
    \label{eq:20}
\end{equation}
where $G''_{\mathrm{min}}$ [Pa] is the local minimum of the loss modulus $G''$, and $G'_{\mathrm{min}}$ [Pa] is the storage modulus $G'$ evaluated at the same angular frequency $\omega_{\mathrm{min}}$ [rad s$^{-1}$] at which $G''$ reaches its local minimum rather than the minimum of $G'$ itself. Tan et al. \cite{tan2022quantitative,tan2021determining} subsequently proposed a modified scaling law expressed as
\begin{equation}
    \frac{G'_{\mathrm{min}}}{G''_{\mathrm{min}}}
    = 0.317\left(\frac{L}{l_{\mathrm{e}}}\right)^{0.82}.
    \label{eq:21}
\end{equation}
In this study, the modified scaling law given by Eq.~\eqref{eq:21} was adopted to calculate the micellar length $L$ as
\begin{equation}
    \begin{aligned}
        L
        &= l_{\mathrm{e}}
           \left(
             \frac{G'_{\mathrm{min}}}{0.317G''_{\mathrm{min}}}
           \right)^{1/0.82} \\
        &\simeq \frac{1}{\left(G_0^{5}k_{\mathrm{B}}T\right)^{1/9}}
           \left(
             \frac{2m}{5N_{\mathrm{A}}w}
             \frac{\sigma_{\mathrm{E}}}{c_{\mathrm{V}}}
           \right)^{2/3}
           \left(
             \frac{G'_{\mathrm{min}}}{0.317G''_{\mathrm{min}}}
           \right)^{1/0.82}.
    \end{aligned}
    \label{eq:22}
\end{equation}

\begin{figure}[t!]
\centering
\includegraphics[width=0.92\columnwidth]{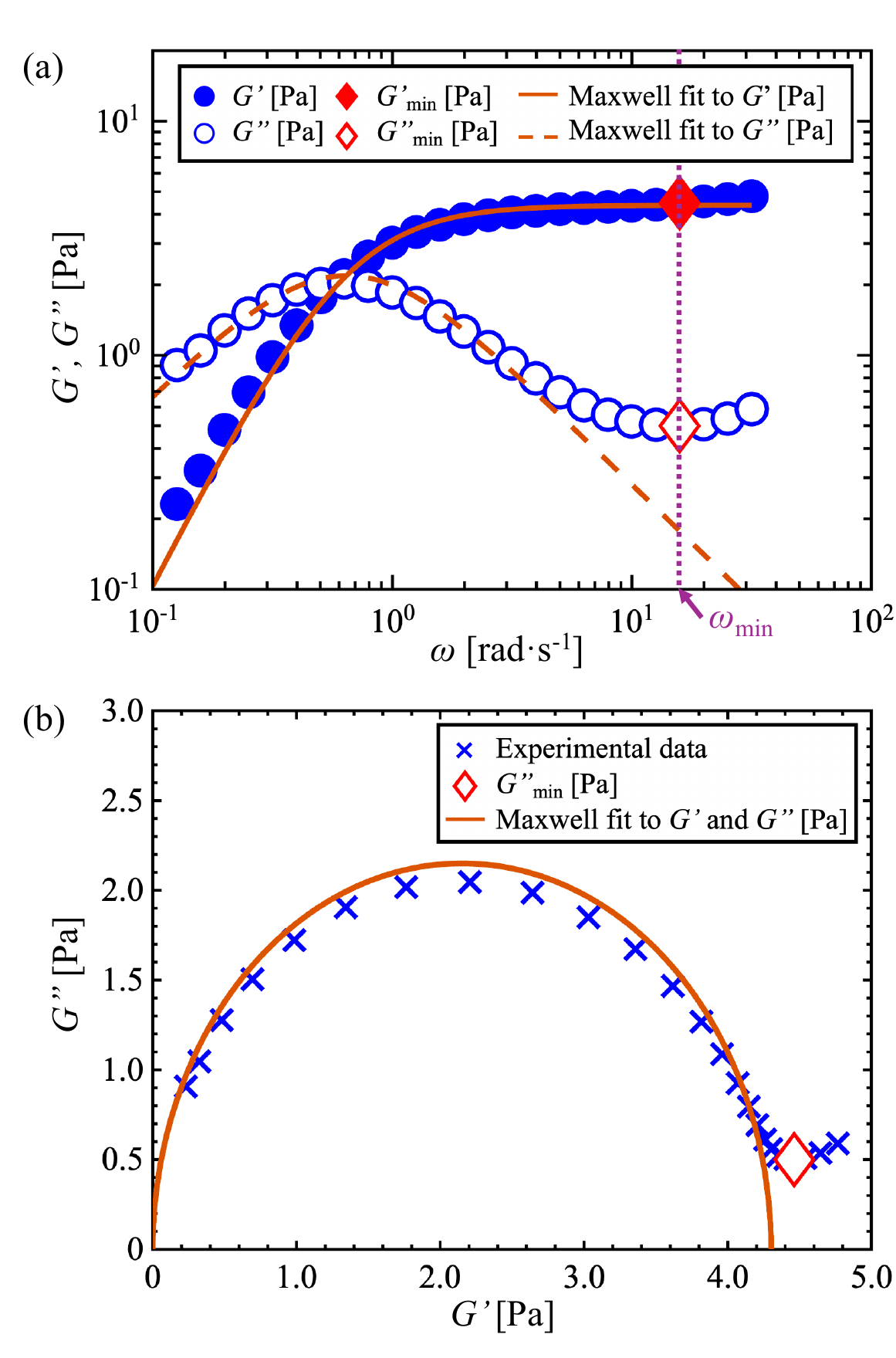}
\caption{Dynamic viscoelasticity of Solution VI measured using a cone-and-plate rotational rheometer at $25\,{}^{\circ}\mathrm{C}$ with a strain amplitude of 0.01 (1\%). (a) Storage modulus $G'$ and loss modulus $G''$ as functions of angular frequency $\omega$. The lines represent single-mode Maxwell fits in the main relaxation region, and the indicated points represent $G'_{\mathrm{min}}$ and $G''_{\mathrm{min}}$. The vertical dashed line indicates the angular frequency $\omega_{\mathrm{min}}$ at which $G''$ reaches its local minimum. (b) Cole--Cole plot of $G''$ versus $G'$.}
\label{fig:9}
\end{figure}

For each sample solution, the storage modulus $G'(\omega)$ and loss modulus $G''(\omega)$ were measured as functions of angular frequency $\omega$ [rad s$^{-1}$] using a cone-and-plate rotational rheometer at $25\,{}^{\circ}\mathrm{C}$ with a strain amplitude of 0.01 ($1\,\%$). Figure~\ref{fig:9} shows representative dynamic viscoelastic measurements for Solution VI.
Figure~\ref{fig:9}(a) shows the frequency dependences of $G'$ and $G''$, including the local minimum $G''_{\mathrm{min}}$ and the corresponding $G'_{\mathrm{min}}$ at $\omega_{\mathrm{min}}$. The linear viscoelastic behavior of CTAB/NaSal wormlike micellar solutions has been reported to follow a single-mode Maxwell model. \cite{inoue2005nonlinear} Accordingly, the plateau modulus $G_{0}$, which is the high-frequency limit of $G'$ in the single-mode Maxwell model, and shear relaxation time $\tau$ [s] were determined by fitting the measured $G'(\omega)$ and $G''(\omega)$ in the main relaxation region to
\begin{equation}
    G'(\omega)
    = G_0 \frac{(\omega\tau)^2}{1+(\omega\tau)^2},
    \label{eq:23}
\end{equation}
\begin{equation}
    G''(\omega)
    = G_0 \frac{\omega\tau}{1+(\omega\tau)^2}.
    \label{eq:24}
\end{equation}
The measured moduli generally agreed with the single-mode Maxwell model in the main relaxation region, whereas deviations associated with local wormlike micellar motions were observed at high angular frequencies.
Figure~\ref{fig:9}(b) shows the Cole--Cole plot of $G''$ versus $G'$. For the single-mode Maxwell model, this plot forms a semicircle with diameter $G_{0}$. \cite{pathak2006rheooptics} The measured values approximately followed this semicircular trajectory in the main relaxation region, which supports the evaluation of $G_{0}$ and $\tau$ using the single-mode Maxwell model. The same analysis was performed for all sample solutions.

Table~\ref{tab:3} lists the values of $G_{0}$, $G'_{\mathrm{min}}$, $G''_{\mathrm{min}}$, and $\tau$ obtained from dynamic viscoelastic measurements together with the micellar length $L$ calculated using the modified scaling law in Eq.~\eqref{eq:22}. 
\begin{table}[t!]
    \centering
    \caption{Linear viscoelastic parameters and micellar lengths $L$, estimated using the modified scaling law in Eq.~(22) for Solutions I--XVII. Here, $G_0$ is the plateau modulus, $\tau$ is the shear relaxation time, and $\omega_{\mathrm{min}}$ is the angular frequency at which $G''$ reaches a local minimum. $G'_{\mathrm{min}}$ and $G''_{\mathrm{min}}$ are the storage and loss moduli, respectively, at $\omega_{\mathrm{min}}$. These values were obtained from dynamic viscoelastic measurements performed using a cone-and-plate rotational rheometer at $25\,{}^{\circ}\mathrm{C}$ with a strain amplitude of 0.01 (1\%). The reported variations in $L$ were determined by propagating the standard deviation of the extensional stress $\sigma_{\mathrm{E}}$ in the EC regime through the micellar length calculation. The micellar length $L$ was not determined for Solution VII.}
    \label{tab:3}
    \setlength{\tabcolsep}{3pt}
    \renewcommand{\arraystretch}{1.35}
    \begin{tabular*}{\columnwidth}{@{\extracolsep{\fill}}lccccc@{}}
    \hline\hline
    Solution
    & $G_0$
    & $G'_{\mathrm{min}}$
    & $G''_{\mathrm{min}}$
    & $\tau$
    & $L$ \\
    No.
    & [Pa]
    & [Pa]
    & [Pa]
    & [s]
    & [$\mu\mathrm{m}$] \\
    \hline
    I    & 0.10 & 0.06 & 0.04 & 0.82 & $41.7 \pm 4.1$ \\
    II   & 0.30 & 0.30 & 0.39 & 1.32 & $26.5 \pm 1.0$ \\
    III  & 0.93 & 0.60 & 0.37 & 1.93 & $18.8 \pm 1.3$ \\
    IV   & 1.64 & 1.64 & 0.28 & 2.61 & $14.6 \pm 1.1$ \\
    V    & 3.37 & 3.39 & 0.42 & 2.39 & $12.4 \pm 2.1$ \\
    VI   & 4.28 & 4.46 & 0.44 & 1.56 & $11.7 \pm 2.0$ \\
    VII  & 3.23 & --- & --- & --- & --- \\
    VIII & 4.13 & 4.11 & 0.46 & 0.86 & $6.7 \pm 0.1$ \\
    IX   & 4.19 & 4.22 & 0.42 & 1.67 & $9.5 \pm 1.0$ \\
    X    & 4.58 & 4.39 & 0.36 & 2.59 & $8.6 \pm 0.9$ \\
    XI   & 3.82 & 3.84 & 0.49 & 3.29 & $12.5 \pm 2.5$ \\
    XII  & 3.82 & 3.86 & 0.50 & 3.28 & $11.8 \pm 1.5$ \\
    XIII & 4.01 & 4.68 & 0.43 & 2.21 & $12.9 \pm 2.8$ \\
    XIV  & 3.10 & 3.39 & 0.52 & 0.67 & $8.7 \pm 1.4$ \\
    XV   & 2.99 & 2.83 & 0.34 & 0.54 & $6.9 \pm 1.8$ \\
    XVI  & 3.48 & 3.38 & 0.73 & 0.39 & $8.8 \pm 0.9$ \\
    XVII & 3.97 & 2.48 & 0.57 & 0.22 & $3.9 \pm 0.5$ \\
    \hline\hline
    \end{tabular*}
\end{table}
The micellar length $L$ was calculated using the mean extensional stress because the extensional stress $\sigma_{\mathrm{E}}$ exhibited slight variations within the EC regime. The variation in $L$ was evaluated by propagating the standard deviation of $\sigma_{\mathrm{E}}$ through the entanglement length $l_{\mathrm{e}}$ using Eq.~\eqref{eq:22}. For Solution VII, however, no intersection of $G'$ and $G''$ was observed within the measured angular-frequency range, which prevented a determination of $\tau$ using the same criterion as for the other samples. Furthermore, evaluating $G'_{\mathrm{min}}$ and $G''_{\mathrm{min}}$ as required for the modified scaling law was relatively challenging because the local minimum of $G''$ could not be clearly identified. Consequently, $L$ was not calculated for Solution VII, and this solution was excluded from subsequent correlation analyses involving $L$.

The consistency of the micellar length estimates with previous results was examined by comparison with Granek and Cates. \cite{granek1992stress} Dynamic viscoelastic measurements were performed for a CTAB/NaSal aqueous solution with the same composition as in their study ($0.006\,\mathrm{M}$ CTAB and $0.012\,\mathrm{M}$ NaSal), which was prepared separately from Solutions I--XVII listed in Table~\ref{tab:1}. The correlation length $\xi$ and entanglement length $l_{\mathrm{e}}$ obtained in this study were $313\,\mathrm{nm}$ and $3.05 \pm 0.20\,\mu\mathrm{m}$, respectively, which are in good agreement with their reported values of approximately $350\,\mathrm{nm}$ and $3\,\mu\mathrm{m}$.
Granek and Cates \cite{granek1992stress} estimated a lower bound of $L > 11\,\mu\mathrm{m}$ using the conventional Cates scaling law. To compare this value with the modified scaling law of Tan et al., the corresponding lower bound of the modulus ratio was calculated using $l_{\mathrm{e}} = 3\,\mu\mathrm{m}$ and Eq.~\eqref{eq:20}, and then converted using Eq.~\eqref{eq:22}, yielding a value of $L > 59\,\mu\mathrm{m}$. In contrast, the micellar lengths obtained in this study were $L = 15.2 \pm 1.0\,\mu\mathrm{m}$ using the conventional scaling law and $L = 114.9 \pm 7.7\,\mu\mathrm{m}$ using the modified scaling law. Both estimates exceeded the corresponding lower bounds of $11\,\mu\mathrm{m}$ and $59\,\mu\mathrm{m}$, respectively.
These results show that the structural parameters estimated in this study are consistent with the previous findings of Granek and Cates, \cite{granek1992stress} which supports the use of the modified scaling law proposed by Tan et al. \cite{tan2022quantitative,tan2021determining} to estimate the micellar length $L$.

The sensitivity of the estimated $L$ to the strain amplitude used in dynamic viscoelastic measurements was examined as a supplementary analysis. The original measurements were performed at a strain amplitude of 0.01 ($1\,\%$) in the linear viscoelastic regime. In contrast, the filament undergoes large deformation during the liquid dripping process, with the extensional strain $\varepsilon$ ranging from approximately $1.9$ to $4.1$ in the EC regime for Solution VI, as shown in Fig.~\ref{fig:5}(c). Additional dynamic viscoelastic measurements were performed for Solution VI at a strain amplitude of 3 ($300\,\%$) to examine the influence of large strain amplitudes on the estimated $L$.
At a strain amplitude of 3, the plateau modulus was $G_0 = 1.70\,\mathrm{Pa}$, while $G'_{\mathrm{min}} = 1.61\,\mathrm{Pa}$ and $G''_{\mathrm{min}} = 0.32\,\mathrm{Pa}$. Substituting these values into Eq.~\eqref{eq:22} yielded an estimated micellar length of $L = 9.8 \pm 1.7\,\mu\mathrm{m}$, comparable to $L = 11.7 \pm 2.0\,\mu\mathrm{m}$ obtained at a strain amplitude of 0.01 (Table~\ref{tab:3}). Although the modified scaling law was developed based on linear viscoelasticity and its applicability to large-amplitude oscillatory shear is not established, this supplementary comparison indicates that the estimated $L$ for Solution VI did not differ substantially between the two strain amplitudes.

Based on these findings, the micellar lengths $L$ estimated from linear dynamic viscoelastic measurements at a strain amplitude of 0.01 were adopted for the subsequent analysis. In the following section, the relationships between $L$ and the stress-optical parameters are examined to investigate how micellar structure contributes to the stress-optical response under uniaxial extensional flow.

\subsection{\label{sec:4.2}Dependence of the stress-optical response on micellar length}
This section examines the relationships between the micellar length $L$ and the intrinsic birefringence, polarizability anisotropies, and stress-optical coefficient to investigate the influence of micellar length on the stress-optical response under uniaxial extensional flow.

\begin{figure}[t!]
\centering
\includegraphics[width=0.96\columnwidth]{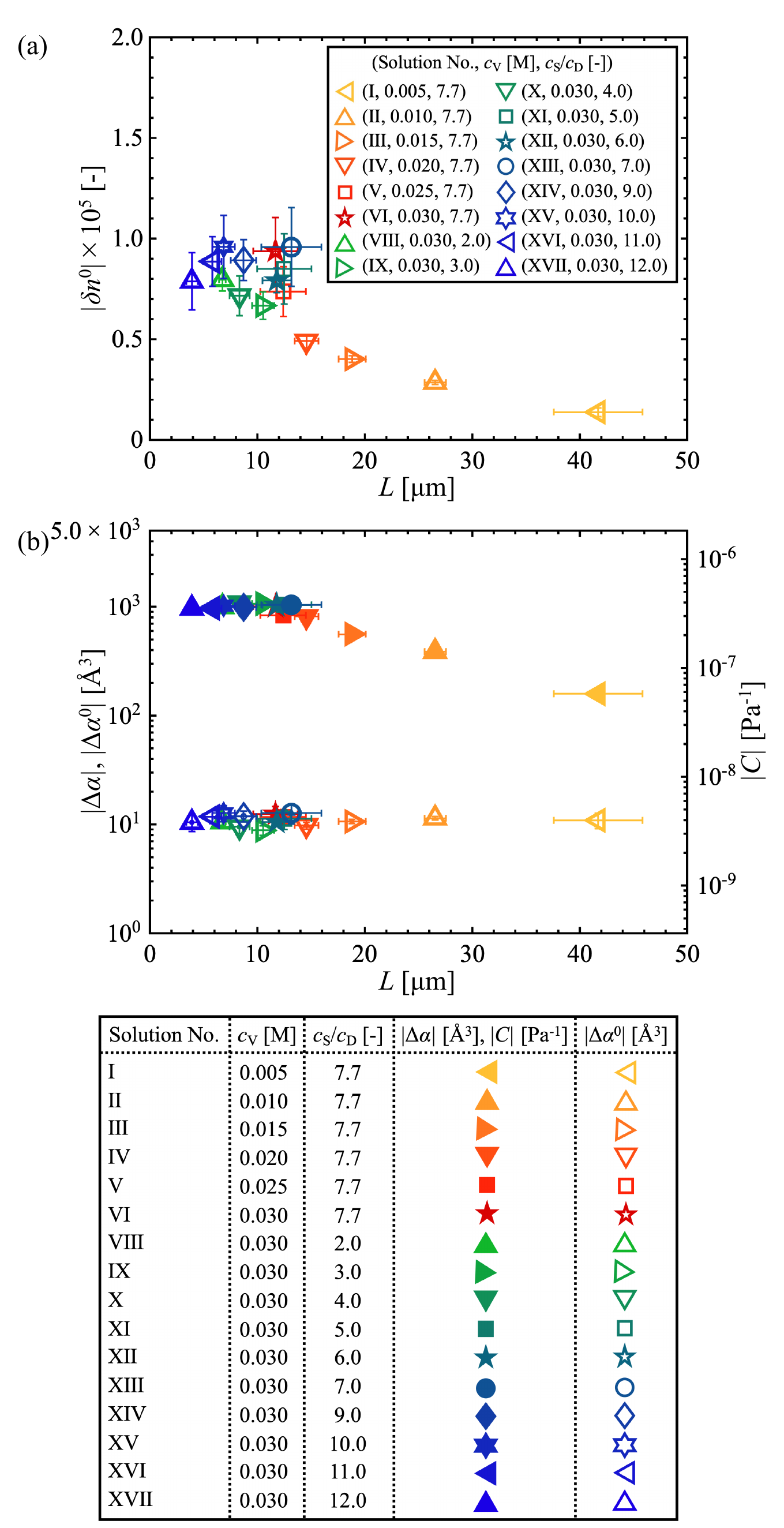}
\caption{
Dependence of the rheo-optical parameters on the micellar length $L$. (a) Magnitude of the intrinsic birefringence $|\delta n^{0}|$. (b) Magnitudes of the polarizability anisotropies $|\Delta\alpha|$ and $|\Delta\alpha^{0}|$ and the stress-optical coefficient $|C|$. In (b), the polarizability anisotropies are plotted on the left vertical axis and $|C|$ on the right vertical axis, with the scales related by the proportionality between $|\Delta\alpha|$ and $|C|$. Open symbols represent $|\Delta\alpha^{0}|$, whereas filled symbols represent $|\Delta\alpha|$ and $|C|$. The legend for panel (b) is placed below the panel. Solution VII was excluded from the analysis.}
\label{fig:10}
\end{figure}

Figure~\ref{fig:10} shows the relationships between the micellar length $L$ and the stress-optical parameters. 
Figure~\ref{fig:10}(a) shows the magnitude of the intrinsic birefringence $|\delta n^{0}|$, whereas Fig.~\ref{fig:10}(b) shows the magnitudes of the polarizability anisotropies $|\Delta\alpha|$ and $|\Delta\alpha^{0}|$ together with the stress-optical coefficient $|C|$.
Figure~\ref{fig:10}(a) shows that $|\delta n^{0}|$ decreased with increasing $L$ for Solutions I--VI at a fixed $c_{\mathrm{S}}/c_{\mathrm{D}} = 7.7$. In these solutions, $L$ decreased with increasing $c_{\mathrm{V}}$, whereas $|\delta n^{0}|$ increased, as shown in Fig.~\ref{fig:7}(a). Thus, the observed negative correlation between $|\delta n^{0}|$ and $L$ may primarily reflect changes in $c_{\mathrm{V}}$ rather than a direct dependence on $L$. Consistent with this interpretation, no clear systematic dependence of $|\delta n^{0}|$ on $L$ was observed for Solutions VIII--XVII at a fixed $c_{\mathrm{V}} = 0.030\,\mathrm{M}$ despite variations in $L$ from approximately $3.9$ to $13.2\,\mu\mathrm{m}$.

Next, we consider the polarizability anisotropy of equivalent Kuhn segments $|\Delta\alpha|$ and the stress-optical coefficient $|C|$. As shown in Fig.~\ref{fig:10}(b), these quantities exhibited corresponding changes with the micellar length $L$ because they are proportional according to Eq.~\eqref{eq:13}, with the temperature and mean refractive index fixed in this study. Consequently, their data points overlap when plotted against their respective vertical axes.
For Solutions I--VI, at a fixed $c_{\mathrm{S}}/c_{\mathrm{D}} = 7.7$, both $|\Delta\alpha|$ and $|C|$ decreased with increasing $L$. In these solutions, $L$ decreased with increasing $c_{\mathrm{V}}$, whereas $|\Delta\alpha|$ and $|C|$ increased, as shown in Fig.~\ref{fig:8}(a). Thus, the observed negative correlations with $L$ may primarily reflect changes in $c_{\mathrm{V}}$ rather than a direct dependence on $L$. Consistent with this interpretation, no clear systematic dependence of $|\Delta\alpha|$ or $|C|$ on $L$ was observed for Solutions VIII--XVII at a fixed $c_{\mathrm{V}} = 0.030\,\mathrm{M}$.
The polarizability anisotropy of micellar segments $|\Delta\alpha^{0}|$, represented by the open symbols in Fig.~\ref{fig:10}(b), ranged from $7.9$ to $12.7$~\AA$^{3}$, with no clear systematic dependence on $L$. Since $|\Delta\alpha^{0}|$ characterizes the local optical anisotropy of individual micellar segments, these results suggest that their optical anisotropy is relatively insensitive to changes in micellar length.

\begin{figure}[b!]
\centering
\includegraphics[width=0.98\columnwidth]{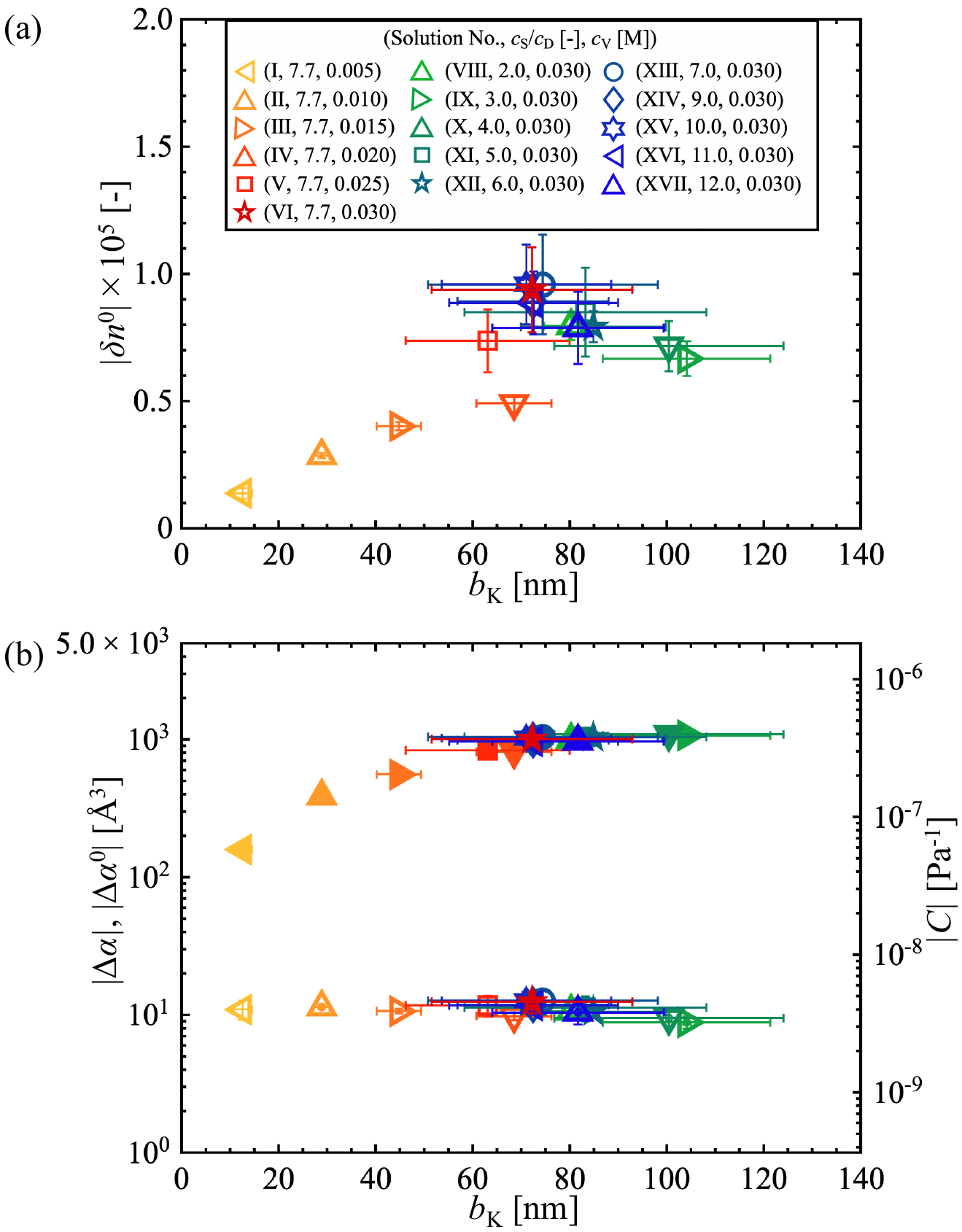}
\caption{Dependence of the rheo-optical parameters on the Kuhn length $b_{\mathrm{K}}$. (a) Magnitude of the intrinsic birefringence $|\delta n_0|$. (b) Magnitudes of the polarizability anisotropies $|\Delta\alpha|$ and $|\Delta\alpha^0|$ and the stress-optical coefficient $|C|$. In (b), the polarizability anisotropies are plotted on the left vertical axis and $|C|$ on the right vertical axis, with the scales related by the proportionality between $|\Delta\alpha|$ and $|C|$. The legend is omitted because it is identical to that in Fig.~\ref{fig:10}(b). Solution VII was excluded from the analysis.}
\label{fig:11}
\end{figure}

Furthermore, the dependence of the stress-optical response on the Kuhn length $b_{\mathrm{K}}$ was examined because the polarizability anisotropy $\Delta\alpha$ characterizes the optical anisotropy of an equivalent Kuhn segment. Figure~\ref{fig:11} shows the relationships between $b_{\mathrm{K}}$ and the intrinsic birefringence, polarizability anisotropies, and stress-optical coefficient. 
For Solutions I--VI, at a fixed $c_{\mathrm{S}}/c_{\mathrm{D}} = 7.7$, $|\delta n^{0}|$, $|\Delta\alpha|$, and $|C|$ tended to increase with increasing $b_{\mathrm{K}}$. In contrast, for Solutions VIII--XVII at a fixed $c_{\mathrm{V}} = 0.030\,\mathrm{M}$, no clear systematic dependence of these quantities on $b_{\mathrm{K}}$ was observed. The polarizability anisotropy of micellar segments $|\Delta\alpha^{0}|$ remained approximately constant regardless of $b_{\mathrm{K}}$ for both groups.
The value of $b_{\mathrm{K}}$ is calculated from the ratio $|\Delta\alpha|/|\Delta\alpha^{0}|$ according to Eq.~\eqref{eq:17}, while $|\Delta\alpha|$ is proportional to $|C|$ according to Eq.~\eqref{eq:13}. Thus, the positive correlations between $b_{\mathrm{K}}$ and $|\Delta\alpha|$ or $|C|$ partly arise from their mathematical relationships and do not necessarily indicate a direct effect of Kuhn length on the stress-optical response. Furthermore, the observed trends may reflect an effect of concentration given that $c_{\mathrm{V}}$ also varied among Solutions I--VI. These results suggest that the Kuhn length $b_{\mathrm{K}}$ alone is insufficient to explain the differences in the stress-optical response among the samples under the conditions investigated.

The observed correlations between the micellar length $L$ or Kuhn length $b_{\mathrm{K}}$ and the stress-optical response may reflect changes in the micellar molar concentration $c_{\mathrm{V}}$ and the mathematical relationships among the evaluated parameters. In contrast, no clear systematic dependence of the stress-optical response on $L$ or $b_{\mathrm{K}}$ was observed at a fixed $c_{\mathrm{V}}$. These results suggest that the stress-optical response under uniaxial extensional flow is more strongly associated with $c_{\mathrm{V}}$ and hence the number density of micellar segments than with the micellar length or flexibility under the conditions investigated in this work.

\section{\label{sec:5}Discussion of the stress-optical response based on the internal structure of wormlike micelles}
\subsection{\label{sec:5.1}Key factors governing the stress-optical response}
\begin{figure}[t!]
\centering
\includegraphics[width=0.96\columnwidth]{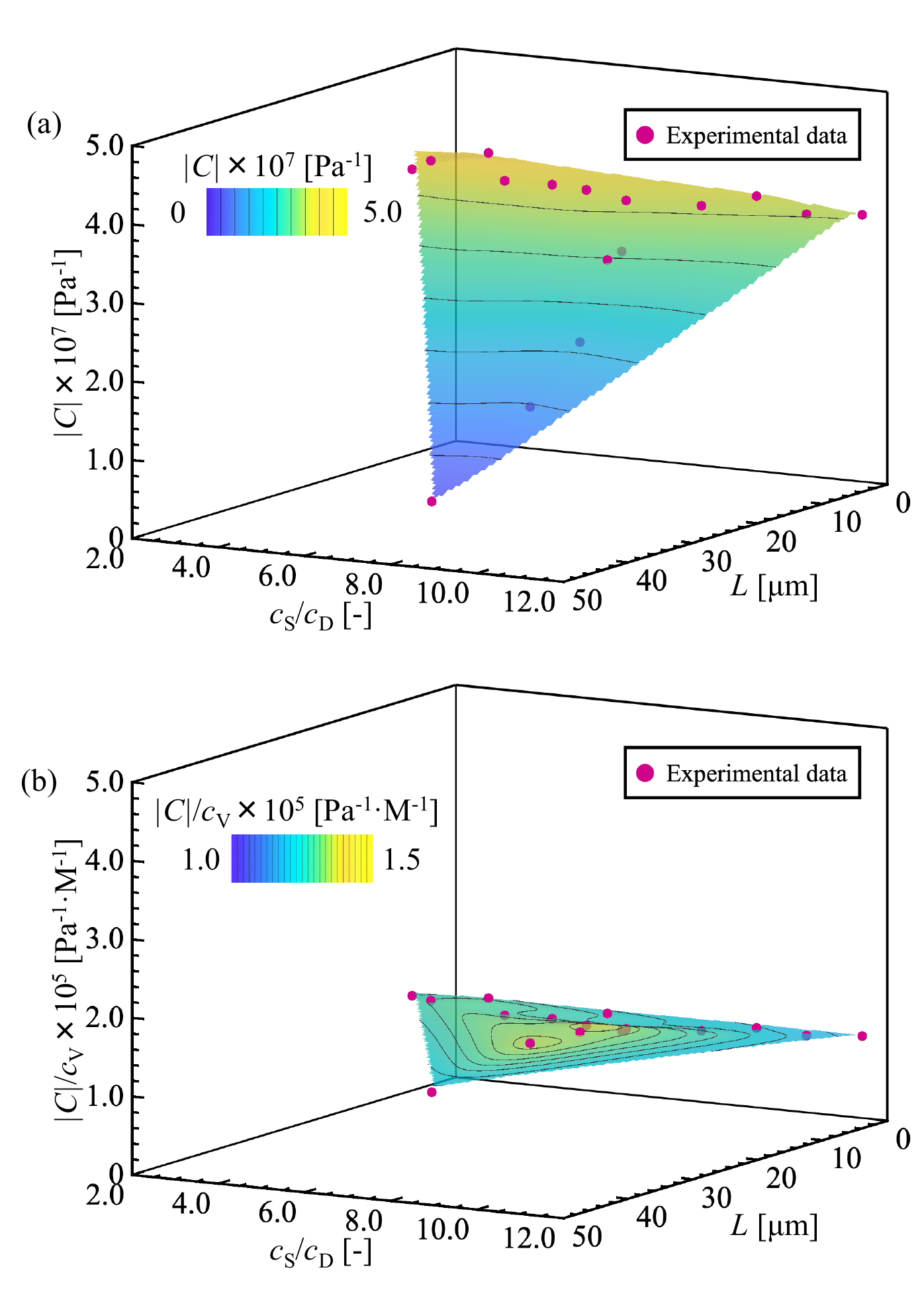}
\caption{
Three-dimensional distributions of (a) the magnitude of the stress-optical coefficient $|C|$ and (b) its normalized value $|C|/c_{\mathrm{V}}$ as functions of the CTAB/NaSal molar ratio $c_{\mathrm{S}}/c_{\mathrm{D}}$ and micellar length $L$. Magenta symbols represent experimental data, while colored surfaces and contours are visual guides obtained by interpolation and partial extrapolation. The contour intervals are $0.5 \times 10^{-7}\,\mathrm{Pa}^{-1}$ in (a) and $2.0 \times 10^{-7}\,\mathrm{Pa}^{-1}\,\mathrm{M}^{-1}$ in (b). Solution VII was excluded from the analysis.}
\label{fig:12}
\end{figure}

To further examine the relative importance of the micellar molar concentration $c_{\mathrm{V}}$, CTAB/NaSal molar ratio $c_{\mathrm{S}}/c_{\mathrm{D}}$, and micellar length $L$ in the stress-optical response, their relationships with the stress-optical coefficient $C$ were examined using three-dimensional representations. Figure~\ref{fig:12}(a) shows the distribution of $|C|$ as a function of $c_{\mathrm{S}}/c_{\mathrm{D}}$ and $L$.
The magenta symbols indicate the experimental values, while the colored surface and contours visualize their distribution through interpolation and partial extrapolation. As shown in Fig.~\ref{fig:12}(a), $|C|$ ranged from $0.58 \times 10^{-7}$ to $3.99 \times 10^{-7}\,\mathrm{Pa}^{-1}$ with substantial variation among the samples. However, as discussed in Sec.~\ref{sec:4.2}, the negative correlation between $|C|$ and $L$ may primarily reflect changes in $c_{\mathrm{V}}$ rather than a direct dependence on $L$. Thus, the observed variation in $|C|$ cannot be fully explained by $c_{\mathrm{S}}/c_{\mathrm{D}}$ and $L$ alone, which highlights the need to account for the influence of $c_{\mathrm{V}}$.

The value of $|C|$ was normalized by $c_{\mathrm{V}}$ to account for the effect of the micellar molar concentration $c_{\mathrm{V}}$. As shown in Fig.~\ref{fig:12}(b), $|C|/c_{\mathrm{V}}$ [$\mathrm{Pa}^{-1}\,\mathrm{M}^{-1}$] exhibited a relatively flat distribution with respect to $c_{\mathrm{S}}/c_{\mathrm{D}}$ and $L$, ranging from $(1.16$--$1.49) \times 10^{-5}\,\mathrm{Pa}^{-1}\,\mathrm{M}^{-1}$. The sample-to-sample variation was evaluated relative to the median for the 16 samples excluding Solution VII. The magnitude $|C|$ varied from approximately $-84$ to $+10\,\%$ relative to its median of $3.62 \times 10^{-7}\,\mathrm{Pa}^{-1}$, whereas $|C|/c_{\mathrm{V}}$ varied from approximately $-7$ to $+19\,\%$ relative to its median of $1.25 \times 10^{-5}\,\mathrm{Pa}^{-1}\,\mathrm{M}^{-1}$. Thus, normalization by $c_{\mathrm{V}}$ substantially reduced the sample-to-sample variation. These results suggest that $|C|$ varies approximately in proportion to $c_{\mathrm{V}}$ under the conditions investigated, with no clear systematic dependence on $c_{\mathrm{S}}/c_{\mathrm{D}}$ or $L$ after normalization.

Taken together, these findings suggest that the dependence of the stress-optical coefficient $C$ on concentration is more closely associated with the number density of micellar segments than with changes in the local optical anisotropy of individual micellar segments. As discussed in Sec.~\ref{sec:3.3}, the polarizability anisotropy of individual micellar segments $|\Delta\alpha^{0}|$ showed no clear systematic dependence on solution composition, whereas the number density of micellar segments increases with $c_{\mathrm{V}}$. An increase in segment number density can enhance birefringence by increasing the number of optically anisotropic segments per unit volume provided that their orientation and local optical properties remain comparable. 

\subsection{\label{sec:5.2}Evaluation of the stress-optical response based on micellar volume fraction}
In the previous section, the normalized stress-optical coefficient $|C|/c_{\mathrm{V}}$ showed relatively small variations despite differences in the CTAB/NaSal molar ratio $c_{\mathrm{S}}/c_{\mathrm{D}}$ and micellar length $L$. To interpret this concentration dependence from a geometrical perspective, the geometrical micellar volume fraction $\varPhi_{\mathrm{m}}$ [\%] which is proportional to $c_{\mathrm{V}}$ is introduced to examine its relationship with $|C|$.

In the micellar segment chain model described in Sec.~\ref{sec:2.1} and illustrated in Fig.~\ref{fig:2}, a wormlike micelle consists of disk-shaped micellar segments connected along the axial direction, each containing $m$ micellar units. For a micellar segment with radius $h$ [m] and thickness $w$ [m], its volume is $\pi h^{2}w$ [$\mathrm{m}^{3}$]. Since the number density of micellar segments is $10^{3}c_{\mathrm{V}}N_{\mathrm{A}}/m$ [$\mathrm{m}^{-3}$], the geometrical micellar volume fraction $\varPhi_{\mathrm{m}}$ [\%] is defined as
\begin{equation}
    \varPhi_{\mathrm{m}}
    = \frac{10^{5}\pi h^{2}N_{\mathrm{A}}w}{m}
      c_{\mathrm{V}}.
    \label{eq:25}
\end{equation}
In this study, $h = 2.4\,\mathrm{nm}$, $w = 0.85\,\mathrm{nm}$, and $m = 18$ were assumed for all samples, following the structural model proposed by Shikata et al. \cite{shikata1994rheooptical} The resulting $\varPhi_{\mathrm{m}}$ ranged from approximately $0.26$ to $1.54\,\%$ for $c_{\mathrm{V}} = 0.005$--$0.030\,\mathrm{M}$. This calculation neglects local structural differences at the ends, branches, and cross-links of wormlike micelles and instead approximates their total volume as the sum of individual micellar segment volumes. Thus, $\varPhi_{\mathrm{m}}$ represents the geometrical volume fraction rather than an effective occupied volume fraction that accounts for the spatial extension of micelles due to bending and entanglement.

\begin{figure}[t!]
\centering
\includegraphics[width=0.96\columnwidth]{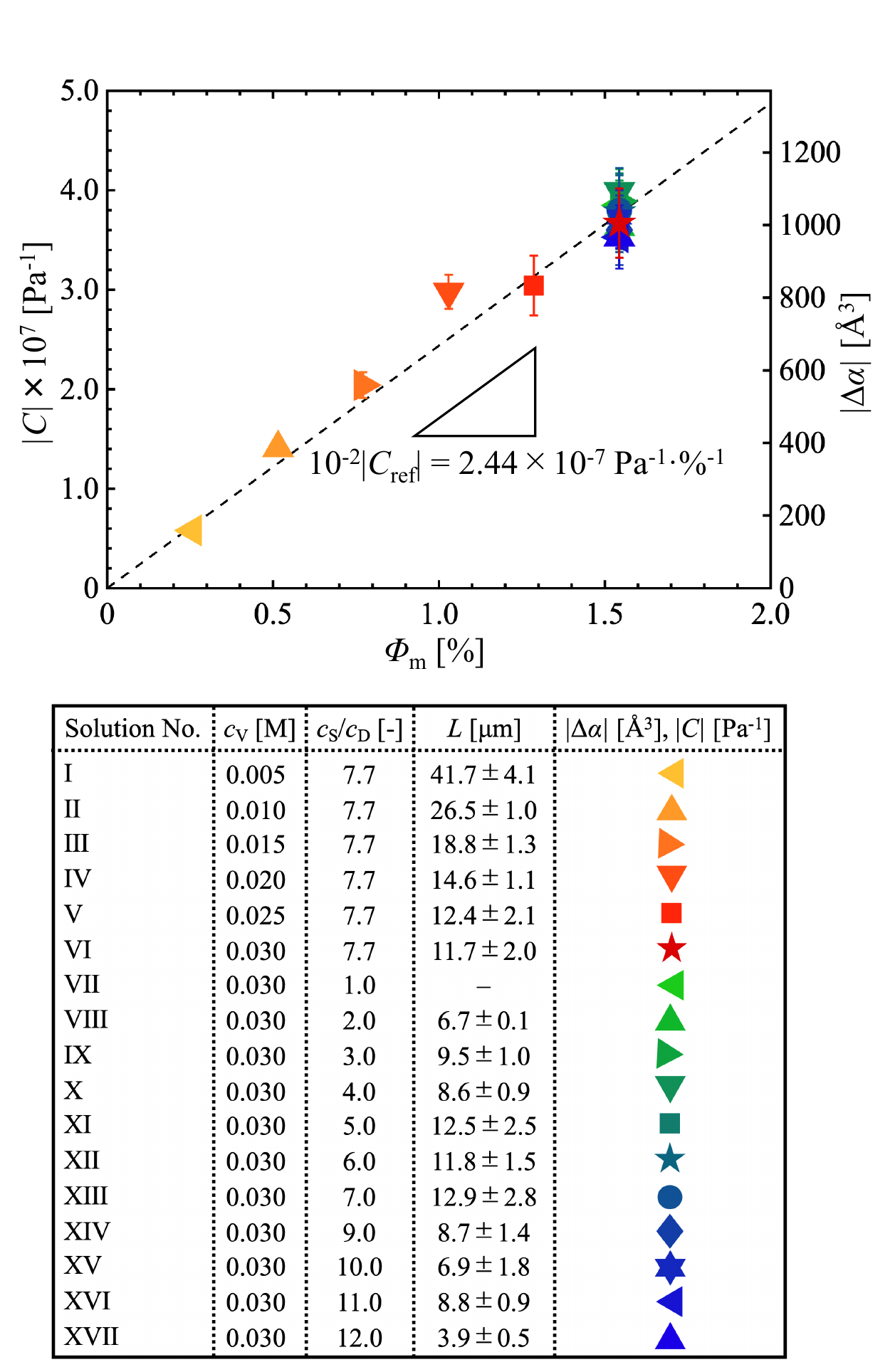}
\caption{
Dependence of the stress-optical coefficient $|C|$ and polarizability anisotropy of equivalent Kuhn segments $|\Delta\alpha|$ on the micellar volume fraction $\varPhi_{\mathrm{m}}$ for Solutions I--XVII. The magnitudes $|C|$ and $|\Delta\alpha|$ are plotted on the left and right vertical axes, respectively, with the scales related by their proportionality in Eq.~\eqref{eq:13}. The dashed line represents a least-squares linear fit of $|C|$ constrained to pass through the origin with a slope of $10^{-2}|C_{\mathrm{ref}}| = 2.44 \times 10^{-7}\,\mathrm{Pa}^{-1}\,\%^{-1}$ and a coefficient of determination of $R^{2} = 0.965$.}
\label{fig:13}
\end{figure}

Figure~\ref{fig:13} shows the relationship between the micellar volume fraction $\varPhi_{\mathrm{m}}$ and the stress-optical response.
As shown in Fig.~\ref{fig:13}, both $|C|$ and $|\Delta\alpha|$ increased with increasing $\varPhi_{\mathrm{m}}$. Because $|\Delta\alpha|$ is proportional to $|C|$ according to Eq.~\eqref{eq:13}, the two quantities share the same data points on their respective vertical axes. To quantify this dependence, $|C|$ was fitted by a least-squares straight line constrained to pass through the origin, assuming proportionality to $\varPhi_{\mathrm{m}}$. The resulting relationship is expressed as
\begin{equation}
    |C| = 10^{-2}|C_{\mathrm{ref}}|\varPhi_{\mathrm{m}}.
    \label{eq:26}
\end{equation}
Here, the coefficient $|C_{\mathrm{ref}}|$ [$\mathrm{Pa}^{-1}$] corresponds to the proportionality coefficient when $\varPhi_{\mathrm{m}}$ is expressed as a dimensionless fraction. $\varPhi_{\mathrm{m}}$ [$\%$] is expressed in percent, and $10^{-2}|C_{\mathrm{ref}}|$ [$\mathrm{Pa}^{-1}\,\%^{-1}$] represents the slope of the fitting line.

To evaluate the fitting accuracy, the coefficient of determination $R^{2}$ was calculated as
\begin{equation}
    R^{2}
    = 1 -
    \frac{
        \displaystyle\sum_{i=1}^{N}
        \left(
            |C_i| - 10^{-2}|C_{\mathrm{ref}}|\varPhi_{\mathrm{m},i}
        \right)^{2}
    }{
        \displaystyle\sum_{i=1}^{N}
        \left(
            |C_i| - \overline{|C|}
        \right)^{2}
    },
    \label{eq:27}
\end{equation}
where $N$ is the number of samples, $|C_i|$ and $\varPhi_{\mathrm{m},i}$ are the magnitude of the stress-optical coefficient and micellar volume fraction for sample $i$, respectively, and $\overline{|C|}$ is the mean of $|C|$ across all samples. The least-squares fitting yielded a slope of $10^{-2}|C_{\mathrm{ref}}| = 2.44 \times 10^{-7}\,\mathrm{Pa}^{-1}\,\%^{-1}$ and $R^{2} = 0.965$.
If the entire solution volume is assumed to be occupied by wormlike micellar material ($\varPhi_{\mathrm{m}} = 100\,\%$), Eq.~\eqref{eq:26} yields $|C| = |C_{\mathrm{ref}}| = 2.44 \times 10^{-5}\,\mathrm{Pa}^{-1}$. However, this value represents a formal extrapolation beyond the investigated range rather than an experimentally verified stress-optical coefficient.
According to Eq.~\eqref{eq:26}, the ratio $|C|/\varPhi_{\mathrm{m}}$ corresponds to the slope $10^{-2}|C_{\mathrm{ref}}|$ of the fitted line. To examine the sample-to-sample variation in this relationship, $|C|/\varPhi_{\mathrm{m}}$ was calculated individually for each sample using the measured $|C|$ and calculated $\varPhi_{\mathrm{m}}$. The resulting values fell in the range of $(2.25$--$2.90) \times 10^{-7}\,\mathrm{Pa}^{-1}\,\%^{-1}$, corresponding to approximately $-8$ to $+18\,\%$ relative to the median for the 17 samples. This relatively small variation suggests that $|C|$ can be approximately described using a common proportionality coefficient $10^{-2}|C_{\mathrm{ref}}|$ under the conditions investigated, despite differences in the CTAB/NaSal molar ratio and micellar length.

One important finding of this study is that the magnitude of the stress-optical coefficient $|C|$ can be described by an approximately linear relationship with the micellar volume fraction $\varPhi_{\mathrm{m}}$ despite substantial differences in solution composition and micellar structure. Specifically, the micellar molar concentration $c_{\mathrm{V}}$, CTAB/NaSal molar ratio $c_{\mathrm{S}}/c_{\mathrm{D}}$, and micellar length $L$ varied by factors of up to 6, 12, and approximately 11, respectively. Nevertheless, the values of $|C|/\varPhi_{\mathrm{m}}$ remained within $-8$ to $+18\,\%$ relative to the median. This finding indicates that the stress-optical response can be characterized using a common proportionality coefficient $|C_{\mathrm{ref}}|$.
Notably, the approximately linear relationship between $|C|$ and $\varPhi_{\mathrm{m}}$ was observed across solutions potentially containing entangled wormlike micelles and branched or networked structures. This suggests that the micellar volume fraction provides a common basis to characterize the stress-optical response even when the internal micellar morphology differs among samples. Although $\varPhi_{\mathrm{m}}$ is proportional to $c_{\mathrm{V}}$ under the present structural model, expressing the concentration dependence in terms of $\varPhi_{\mathrm{m}}$ provides a geometrical interpretation and enables $|C|$ to be estimated using Eq.~\eqref{eq:26} with a common empirical proportionality coefficient $|C_{\mathrm{ref}}|$.
This empirical relationship may reduce the need for individual calibration of the stress-optical coefficient for each solution composition and thus facilitate quantitative extensional stress evaluation from birefringence measurements. Further validation of its applicability and the accuracy of the resulting estimations remains necessary, particularly under different flow kinematics such as shear flow.

\section{\label{sec:6}Conclusions}
In this study, the relationships between the stress-optical response and micellar structure under uniaxial extensional flow were investigated for 17 CTAB/NaSal aqueous solutions with systematically varied CTAB/NaSal molar ratios $c_{\mathrm{S}}/c_{\mathrm{D}}$ and micellar molar concentrations $c_{\mathrm{V}}$. Full-field extensional rheo-optical data and dynamic viscoelastic measurements were analyzed to evaluate the intrinsic birefringence, stress-optical coefficient, polarizability anisotropies of micellar segments and equivalent Kuhn segments, and micellar length.

Analysis of the full-field extensional rheo-optical data showed that the optical principal-axis direction was nearly aligned with the extensional direction in the elasto-capillary regime, and the birefringence was approximately proportional to the extensional stress, consistent with the stress-optical rule. The rheo-optical data were further used to evaluate the intrinsic birefringence and the polarizability anisotropies of micellar segments and equivalent Kuhn segments. The magnitudes of the intrinsic birefringence $|\delta n^{0}|$, stress-optical coefficient $|C|$, and polarizability anisotropy of equivalent Kuhn segments $|\Delta\alpha|$ increased with increasing $c_{\mathrm{V}}$, while showing no clear systematic dependence on $c_{\mathrm{S}}/c_{\mathrm{D}}$ at fixed $c_{\mathrm{V}}$. In contrast, the polarizability anisotropy of individual micellar segments $|\Delta\alpha^{0}|$ showed no clear systematic dependence on solution composition, which suggests that their local optical anisotropy remained approximately constant. These findings suggest that variations in the stress-optical coefficient $C$ are more closely associated with the number density of micellar segments than with changes in the local optical anisotropy of individual segments.

The micellar length $L$ was estimated from dynamic viscoelastic measurements using the modified scaling law in Eq.~(22), while the Kuhn length $b_{\mathrm{K}}$ was evaluated from the polarizability anisotropies. For samples with varying $c_{\mathrm{V}}$, $|\delta n^{0}|$ and $|C|$ showed negative correlations with $L$, whereas $|C|$ showed a positive correlation with $b_{\mathrm{K}}$. However, these correlations were not clearly observed at fixed $c_{\mathrm{V}}$. The negative correlations with $L$ may primarily reflect changes in $c_{\mathrm{V}}$, while the positive correlation with $b_{\mathrm{K}}$ partly arises from its mathematical relationship with $|C|$. Thus, neither $L$ nor $b_{\mathrm{K}}$ alone adequately explains the differences in the stress-optical response. Normalization of $|C|$ by $c_{\mathrm{V}}$ substantially reduced the sample-to-sample variation.

An important finding of this study is that $|C|$ could be approximately described by a single proportional relationship with the geometrical micellar volume fraction $\varPhi_{\mathrm{m}}$ within the range of solution compositions investigated, despite the absence of a clear systematic dependence on the micellar length $L$ at fixed $c_{\mathrm{V}}$. Although $c_{\mathrm{V}}$, $c_{\mathrm{S}}/c_{\mathrm{D}}$, and $L$ varied by factors of up to 6, 12, and approximately 11, respectively, the values of $|C|/\varPhi_{\mathrm{m}}$ remained within approximately $-8$ to $+18\,\%$ relative to the median. The relationship was characterized by $|C_{\mathrm{ref}}| = 2.44 \times 10^{-5}\,\mathrm{Pa}^{-1}$. This proportionality was observed across samples potentially containing entangled wormlike micelles and branched or networked structures, which suggests that the stress-optical response is more closely associated with the volume fraction of micellar material than with micellar length or morphology alone.

The empirical relationship provides a basis to estimate the stress-optical coefficient $C$ from the micellar volume fraction $\varPhi_{\mathrm{m}}$. This finding could potentially reduce the need for individual calibration and facilitate quantitative extensional stress evaluation from birefringence measurements. However, this relationship does not independently establish the microscopic mechanism governing the concentration dependence of $C$ because $\varPhi_{\mathrm{m}}$ is proportional to $c_{\mathrm{V}}$ under the present structural model. Further investigations are needed to clarify the contributions of micellar flexibility, branching, and cross-linking and to validate the applicability and accuracy of the proposed relationship under different flow kinematics, including shear flow.

\end{CJK}

\begin{acknowledgments}
The present study was supported by the Japan Society for the Promotion of Science, KAKENHI Grant Nos. 23H01343 and was partially based on the results obtained from project JPNP20004, subsidized by the New Energy and Industrial Technology Development Organization (NEDO).
\end{acknowledgments}

\section*{Data Availability Statement}
The data that support the findings of the present study are available from the corresponding author upon reasonable request.

\bibliography{aipsamp}

\end{document}